\documentclass[12pt]{article}
\pdfoutput=1
\usepackage[utf8]{inputenc}
\usepackage{amsfonts,amsmath}
\usepackage{jheppub}

\usepackage{color}
\usepackage{tabularx}

\usepackage{ifthen}
\usepackage{enumerate}
\usepackage{amsmath}
\usepackage{amssymb}
\usepackage[mathscr]{eucal}
\usepackage[errorshow]{tracefnt}
\usepackage{amsmath}
\usepackage{slashed}
\usepackage{amssymb}
\usepackage{array}
\usepackage{amsthm}
\usepackage{eucal}
\usepackage{epsf}
\usepackage{epsfig}
\usepackage{psfrag}
\usepackage{multirow}
\usepackage{wasysym} 
\usepackage{tikz-cd} 
\usepgfmodule{shapes}
\usetikzlibrary{backgrounds, arrows,calc,shapes,decorations.pathreplacing, decorations.markings, automata,positioning}

\newcommand{\nn}{\nonumber}
\newcommand{\rf}{r_{\phi}}

\newcommand{\beqa}{\begin{eqnarray}}
\newcommand{\eeqa}{\end{eqnarray}}

\newcommand{\beq}{\begin{equation}}
\newcommand{\eeq}{\end{equation}}
\newcommand{\bea}{\begin{eqnarray}}
\newcommand{\eea}{\end{eqnarray}}
\newcommand{\ba}{\begin{array}}
\newcommand{\ea}{\end{array}}

\newcommand{\CF}{{\mathcal F}}

\newcommand{\M}{{\mathfrak M}}
\newcommand{\CN}{{\mathcal N}}
\newcommand{\CO}{{\mathcal O}}

\newcommand{\CW}{{\mathcal W}}
\newcommand{\CZ}{{\mathcal Z}}

\newcommand\qt{\tilde q}
\newcommand\bt{\tilde b}
\newcommand\pt{\tilde p}
\newcommand\Qt{\tilde Q}

\newcommand{\be}{\begin{equation}}
\newcommand{\ee}{\end{equation}}

\newcommand{\bpic}{\begin{tikzpicture}}
\newcommand{\epic}{\end{tikzpicture}}

\def\+{{+\!\!\!+}}

\def\a{\alpha} 
\def\b{\beta}

\def\0{\nonumber}

\def\1{{\bf 1}}

\title{A tale of exceptional $3d$ dualities}

\preprint{SISSA 37/2018/FISI}

\author[1,2]{Sergio Benvenuti}

\affiliation[1]{International School of Advanced Studies (SISSA), Via Bonomea 265, 34136 Trieste, Italy}
\affiliation[2]{INFN, Sezione di Trieste, Via Valerio 2, 34127 Trieste, Italy}
\emailAdd{benve79@gmail.com}

\abstract{We consider interesting Seiberg dualities for $Usp$ gauge theories with an antisymmetric, $8$ fundamentals and no superpotential. We reduce to three dimensions and systematically analyze deformations triggered by real and complex masses, reaching a plethora of $\mathcal{N}\!=\!2$ dualities for $U(N)$ and $Usp(2N)$ gauge theories, possibly with monopole superpotentials and Chern-Simons interactions. Special cases of these "exceptional dualities" are: supersymmetry enhancement dualities, "duality appetizers" and many known dualities relating rank-$1$ gauge groups. The $4d$ $\mathcal{N}\!=\!1$ $Usp$ dualities provide a unified perspective on many curious phenomena of $3d$ and $4d$ gauge theories with four supercharges.

Finally, we propose a free mirror for $A_{2N}$ Argyres-Douglas, with its related adjoint-$Usp(2N)$ duality, and we construct a mirror for adjoint-$U(N)$, with an arbitrary number flavors and zero superpotential.}

\begin{document}

\maketitle

\section{Introduction and summary}
Soon after the first Seiberg duality \cite{Seiberg:1994pq}, \cite{Csaki:1996eu} found a very interesting class of $4d$ $\CN\!=\!1$ dualities for $Usp(2N_c)$ with an antisymmetric field and $8$ fundamentals. The set of dual theories was enlarged in \cite{Csaki:1997cu, Spiridonov:2008zr}, and in the case $N_c\!=\!1$ it includes both Seiberg duality \cite{Seiberg:1994pq} and Intriligator-Pouliot duality \cite{Intriligator:1995ne}.

It was recently realized that these $Usp$ theories have a higher dimensional ancestor: the $6d$ $(1,0)$ superconformal field theory called E-string \cite{Kim:2017toz, Razamat:2018gbu,Kim:2018bpg,Kim:2018lfo}. They are also related to domain walls of $5d$ $\CN\!=\!1$ $Usp(2N_c)$ with an antisymmetric field.% and $8$ flavors (recall that $5d$ $\CN\!=\!1$ $Usp(2N_c)$ with antisymmetric and $N_f$ flavors enjoys $E_{N_f+1}$ global symmetry).

In this paper we study the reduction of the $Usp(2N_c) \leftrightarrow Usp(2N_c)$ $4d$ dualities to $3d$. In $3d$ we can turn on real masses, which do not exist in $4d$, this process leads to $3d$ $\CN\!=\!2$ dualities for $Usp(2N_c)$ or $U(N_c)$ gauge theories with a smaller amount of flavors (and possibly monopole superpotentials). As usual, each $3d$ $\CN\!=\!2$ duality implies an integral identity for its $S^3$ supersymmetric partition function. Such $\CZ_{S^3}$'s were studied independently in the mathematical literature \cite{Rains:2003,FOKKO}, where it is also shown that the set of theories is naturally organized in terms of Weyl groups of a sequence of groups $E_7 \rightarrow SO(12) \rightarrow SU(6) \rightarrow \ldots\,\,$. Because of their $6d$ origin with $E_8$ global symmetry, and of the relations with this sequence of groups, we call the dualities of \cite{Csaki:1996eu, Csaki:1997cu, Spiridonov:2008zr} and their $3d$ descendants \emph{exceptional dualities}.

In the case of minimal number of colors for the gauge groups, $N_c=1$, it is possible to flow to various new dualities and also to many known, apparently unrelated, $\CN\!=\!2$ dualities, that appeared scattered in the literature. As far as we know, all studied dualities relating theories with rank-$1$ or rank-$0$ gauge groups are obtained:\footnote{We only allow superpotentials of the type "gauge singlet times meson/monopole".} the mirror of $U(1)$ with $2$ or $1$ flavors \cite{Aharony:1997bx},  Aharony dualities \cite{Aharony:1997gp} for $U(1)$ with $2$ flavors and $Usp(2)\!=\!SU(2)$ with $6$ or $4$ doublets and their real mass deformations \cite{Giveon:2008zn,Willett:2011gp,Benini:2011mf}, the mirror of $U(1)_{\frac{1}{2}}$ with $1$ chiral flavor \cite{Dorey:1999rb}, the dual of $U(1)_{\CW=\M^++\M^-}$, $4$ \cite{Dimofte:2012pd} or $3$ \cite{Benvenuti:2016wet} flavors, the dual of $U(1)_{\CW=\M^+}$, $3$ \cite{Benini:2017dud} or $2$ \cite{Benini:2017dud,Collinucci:2017bwv} flavors, a duality  "$SU(2)_{1}$ with $4$ doublets" $\leftrightarrow$ "$U(1)$ with $2$ flavors" \cite{Amariti:2014lla, Aharony:2014uya} (which explains IR symmetry enhancement in the latter model \cite{Benini:2018bhk}), a duality "$U(1)_{\frac32, \CW=\M^+}$ with $3$ chiral flavors" $\leftrightarrow$ "$U(1)$ with $2$ chiral flavors" \cite{Fazzi:2018rkr} (which explains IR symmetry enhancement in the latter model \cite{Gang:2017lsr, Gang:2018wek,Gaiotto:2018yjh,Benini:2018bhk,Fazzi:2018rkr}).

All these $N_c\!=\!1$ dualities admit a generalization replacing $U(1) \rightarrow U(N_c)$ with adjoint, and $Usp(2) \rightarrow Usp(2N_c)$ with antisymmetric, while keeping the same superpotential and the same number of fundamentals, which we describe in this paper.\footnote{\cite{Willett:2011gp} and \cite{Benini:2011mf} carried out the analog analysis for dualities involving $U(N)_k$ and $Usp(2N)_k$ with arbitrary number of flavors but no rank-$2$ matter.}

For generic $N_c$, it is possible to flow from the dualities of \cite{Csaki:1996eu, Csaki:1997cu, Spiridonov:2008zr} to two other interesting classes of dualities:
\begin{itemize}
\item as pointed out by Aghaei, Amariti and Sekiguchi in \cite{Aghaei:2017xqe}, which provided us the motivation for this paper, one can reach the duality $SU(N_c)$ with adjoint and $1$ flavor $\leftrightarrow$ $\CN=4$ $U(1)$ with $N_c$ flavors. This duality was found in \cite{Benvenuti:2017lle, Benvenuti:2017kud}, where it was derived deforming an $\CN\!=\!4$ mirror symmetry \cite{Intriligator:1996ex,Hanany:1996ie}. It explains the Lagrangians of \cite{Maruyoshi:2016tqk, Maruyoshi:2016aim} for $4d$ $\CN\!=\!2$ Argyres-Douglas theory $(A_1,A_{2N_c-1})$, so the $4d$ supersymmetry enhancements have their origines in the dualities \cite{Csaki:1996eu, Csaki:1997cu, Spiridonov:2008zr}.
\item the "duality appetizer" for $SU(2)_1$ gauge group of \cite{Jafferis:2011ns} was generalized in \cite{Kapustin:2011vz} to $Usp(2N_c)_2$ with antisymmetric and to $U(N_c)_1$ with adjoint. These are dualities between a gauge theory and a set of free fields. They can be reached from the dualities of \cite{Csaki:1996eu, Csaki:1997cu, Spiridonov:2008zr}, turning on complex masses and "axial" real masses, at the end of the process (no flavors left). See also \cite{Amariti:2014lla}. Along the way we find new, similar, infrared free gauge theories with a non zero amount of flavors (section \ref{sec:appetizers}). It looks like the choice of name, "appetizers", was appropriate.
\end{itemize}
Moreover, \cite{Gang:2018huc} recently proposed that $U(1)_{\frac32}$ with $1$ chiral flavor displays supersymmetry enhancement, in section \ref{sec:axial1} we reach a duality which is consistent with the proposal, and naturally leads to conjecture that \emph{$U(N_c)_{\pm \frac32}$ with adjoint and $1$ chiral flavor displays $\CN=2 \rightarrow \CN=4$ supersymmetry enhancement for any $N_c$}.

We hope this convinces that the $4d$ $\CN\!=\!1$ dualities of \cite{Csaki:1996eu, Csaki:1997cu, Spiridonov:2008zr} allow for a unified perspective on many curious phenomena of $3d$ and $4d$ gauge theories with $4$ supercharges. Therefore, we embark in a systematic investigation of the $3d$ $\CN\!=\!2$ dualities descending from \cite{Csaki:1996eu, Csaki:1997cu, Spiridonov:2008zr}.

In section \ref{sec:selfduals} we review the duality web of \cite{Csaki:1996eu, Csaki:1997cu, Spiridonov:2008zr}, and turn on $3d$ "vector-like" real masses to study self-dualities of $U(N_c)$ and $Usp(2N_c)$ theories without Chern-Simons interactions, leaving somewhat more exotic $U(N_c) \leftrightarrow Usp(2N_c)$ and $U(N_c)_{\CW=0} \leftrightarrow U(N_c)_{\CW=\M^+}$ dualities with Chern-Simons interactions to sections \ref{sec:axial1} and \ref{sec:axial2}.

In section \ref{sec:confining} we deform the previous dualities with a complex mass, leading to new "confining" dualities. We work out the full superpotential of the Wess-Zumino duals in each case.

At the end of the paper, we find new dualities which cannot be obtained  from the $4d$ models of \cite{Csaki:1996eu, Csaki:1997cu, Spiridonov:2008zr}, but can be guessed by analogy with the results discussed previously:
\begin{itemize}
\item in section \ref{sec:AD} we argue that the $3d$ mirror of $(A_1,A_{2N})$ AD models is simply $N$ free hypermultiplets. This statement, together with the Lagrangians of \cite{Maruyoshi:2016tqk, Maruyoshi:2016aim}, allows us to discover a $3d$ duality for $Usp(2N_c)$ with adjoint and $2$ flavors. We check numerically the equality of the $S^3$ partition function, but we do not know how to "derive" the duality from a more fundamental one analog to the ones \cite{Csaki:1996eu, Csaki:1997cu, Spiridonov:2008zr}. We nevertheless expect such an uplift to exist.
\item  in section \ref{sec:mirror} we generalize the mirror duality \cite{Intriligator:1996ex} for $U(N_c)$ with adjoint and $2$ flavors, proposing a mirror dual for $U(N_c)$ with adjoint and a generic number $N_f$ flavors, which is a quiver with $N_f\!-\!1$ $U(N_c)$ gauge groups. We use its associated brane setup. This is similar to the  $\CN\!=\!2$ mirror symmetries discussed in \cite{Giacomelli:2017vgk}, for instance for $U(N_c)$ with $N_f$ flavors and no adjoint. These mirror symmetries have the property that the rank of the UV global symmetries is different on the two sides. On the other hand, other complications that arise in \cite{Giacomelli:2017vgk} (non trivial quantum relations in the chiral ring on the quiver side) are absent here. Since it should be possible to turn on a mass for the adjoint and flow to the duality of \cite{Giacomelli:2017vgk}, we hope this will shed light on those issues.
\end{itemize}

\hspace{1cm}\paragraph{Note.} This work overlaps with \cite{AC}. We are grateful to the authors of \cite{AC} for informing us of their results.
%%%%%%%%%%%%%%%%%%%%%%%%%%%%%%%%%%%%%%%%%%%%%%%%%%%%%%%%%%%%%%%
%%%%%%%%%%%%%%%%%%%%%%%%%%%%%%%%%%%%%%%%%%%%%%%%%%%%%%%%%%%%%%%
\section{Vector-like real masses: self-dualities without Chern-Simons}\label{sec:selfduals}
An interesting class of $4d$ $\CN\!=\!1$ dualities \cite{Csaki:1996eu, Csaki:1997cu, Spiridonov:2008zr} states the IR equivalence of two $\CN\!=\!1$ $Usp(2N_c)$ gauge theories with $8$ flavors and an anti-symmetric field. 

We call these dualities self-dualities, even if to be precise they are duality between two different theories, since on the magnetic side there are also gauge singlet fields. These singlet fields appear in the superpotential multiplying mesonics operators, so they can be called "flipping fields".\footnote{In generic theory with $4$ supercharges, "flipping" an operator $\CO$  means adding a gauge singlet field $\b$ and a superpotential term $\delta\CW= \beta\!\cdot\!\CO$. On the other hand, whenever a gauge singlet appears in the superpotential only once and linearly, we can call it "flipping field".} A more precise name is "self-dualities modulo flips".

As explained in \cite{Spiridonov:2008zr, Spiridonov:2009za}, each duality implies an identity for the associated superconformal index, i.e. the $S^3 \times S^1$ partition function. Such integral identities were studied independently in the mathematical literature \cite{Rains:2003}, where it is also pointed out that the set of dual theories is made of $72$ frames, acted upon by the Weyl group of $E_7$.\footnote{See  \cite{Razamat:2017hda} for an $E_8$ action.}

We first describe the $4d$ $Usp \leftrightarrow Usp$ with $8$ flavors dualities (there are $3$ of them, \ref{BASEDUAL}, \ref{BASEDUALseiberg}, \ref{BASEDUALcsaki}). Reducing them to $3d$, one basically obtains the same duality for $3d$ $\CN\!=\!2$ theories. The only difference is that the linear monopole term in the superpotential $\delta \CW=\M_{Usp(2N_c)}$ is generated on both sides of the duality, so the continuos global symmetry is the same in $3d$. Because of the linear monopole superpotential, on both sides of the duality the monopole operators are not in the chiral ring, so the $3d$ chiral ring is isomorphic to the $4d$ chiral ring.

Once we are in $3d$, we can turn on supersymmetric real masses, which do not exist in $4d$, and deform the dualities  to new $3d$ dualities. This process generically breaks both the gauge and the global symmetry and was studied at the level of $S^3$ partition functions $\CZ_{S^3}$ in \cite{FOKKO}. \cite{FOKKO} studied the full set of  dual frames at each level, demonstrating the existence of an action of the Weyl groups of a sequence of Lie groups of decreasing rank: 
\be E_7 \rightarrow SO(12) \rightarrow SU(6) \rightarrow SU(4) \times SU(2) \rightarrow SU(4) \rightarrow \ldots \ee

Instead of going through the whole set of dual theories at each level, we decided to organize the presentation in a different way. In the remaining of this section we describe many dualities that can be obtained from \ref{BASEDUAL}, restricting ourselves to "vector-like" real masses: in this way no Chern-Simons terms are generated and for $U(N_c)$ gauge group the number of fundamentals is equal to the number of anti-fundamentals. We study self-dualities for each of the following five $3d$ $\CN\!=\!2$ models:
\be\label{selfdualscheme}
\begin{tikzpicture}
\node (a) at (0,2) {$Usp(2N_c)_{\CW=\M}$ w/ $8$ flav};
\node (b) at (-2.5,0) {$Usp(2N_c)$ w/ $6$ flav};
\node (c) at (3,0) {$U(N_c)_{\CW=\M^++\M^-}$ w/ $(4,4)$ flav};
\node (d) at (0,-2) {$U(N_c)_{\CW=\M^+}$ w/ $(3,3)$ flav};
\node (e) at (0,-4) {$U(N_c)$ w/ $(2,2)$ flav};
\draw [->,ultra thick, blue] (a) -- (b);\node at (2.9,1) {\tiny{$+,+,+,+,-,-,-,-$}};
\draw [->, ultra thick, red] (a) -- (c);\node at (-2.6,1) {\tiny{$0,0,0,0,0,0,+,-$}};
\draw [->,ultra thick, red] (b) -- (d);\node at (3,-1) {\tiny{$0,0,0,+;0,0,0,-$}};
\draw [->, ultra thick, blue] (c) -- (d);\node at (-2.4,-1) {\tiny{$+,+,+,-,-,-$}};
\draw [->,ultra thick, blue] (d) -- (e);\node at (0.9,-3) {\tiny{$0,0,+;0,0,-$}};
\node (f) at (-5.5,-3) {"Exotic"};
\node (g) at (-5.3,-3.5) {dualities};
\node at (-5,-4) {Sections \ref{sec:axial1},\ref{sec:axial2}};
\draw [->,ultra thick, green] (b) -- (f);
\draw [->,ultra thick, green] (d) -- (g);
\draw [->,ultra thick, green] (e) -- (g);
\end{tikzpicture}\ee
In the scheme above we also indicated the real masses used in the RG flows. Green arrows represent RG flows triggered by "axial" real masses, that will not be discussed in this section. For each of the $5$ models we describe both an \emph{Intriligator-Pouliot like} self-duality (with the maximal number of gauge singlets on the magnetic side) and a \emph{ Csaki et al. like} self-duality, (with a smaller number of gauge singlets on the magnetic side).

\subsection{$Usp(2N_c)$ with $8$ flavors, $\CW=\M$}
We start from an interesting $4d$ duality found more than 20 years ago in \cite{Csaki:1996eu}, which states the IR equivalence of two $\CN\!=\!1$ $Usp(2N_c)$ gauge theories with $8$ flavors and an anti-symmetric field. The antisymmetric field is traceless, with $N_c (2N_c-1) -1$ components, so in the case $N_c=1$ (i.e. $SU(2)$ gauge theory) it is absent. The precise statement of the duality is:
  \be \label{BASEDUAL} \ba{ccc}
\ba{c}Usp(2N_c)\, \text{w/ antisymmetric} \, a \\
 \textrm{and $8$ flavors} \, q_i. \\
   \CW_{4d}=0 \ea 
    &\Longleftrightarrow& 
\ba{c}Usp(2N_c)\, \text{w/ antisymmetric} \, A, \\
 \textrm{ $8$ flavors $Q_i$ and singlets $\mu_{ij;r}$}. \\
   \CW_{4d}=\sum_{r=0}^{N_c-1}\sum_{i<j=1}^8 \mu_{ij; r}tr(Q_iA^r Q_j) \ea 
   \ea \ee
The $\binom{8}{2}=28$ towers of $N_c$ gauge singlet fields $\mu_{ij; r}$ ($i<j=1,\ldots,8$ $r=0,\ldots,N_c-1$) appearing on the r.h.s. are Seiberg mesons flipping the gauge invariant "dressed mesons" $tr(Q_iA^r Q_j)$\footnote{To be precise, we should use the invariant form $\Omega^{\a\b}$ of the $Usp(2N_c)$ group to write the gauge invariants, for instance $tr(Q_i Q_j)$ stands for $\sum_{\a\b}Q_{i \a} \Omega^{\a\b}Q_{j\b}$. We use a shortened notation throughout the paper.} of the magnetic theory. 
   
The global symmetries on both sides of the duality are
\be SU(8)  \times U(1)_{axial} \times U(1)_R \ee
where the $U(8)$ acts on the $8$ flavors, the $U(1)_{axial} = \frac{U(1)_q \times U(1)_a}{U(1)_{anomaly}}$ acts on both the antisymmetric field and the $8$ fundamentals (and in the r.h.s. on the gauge singlets), the $U(1)_{anomaly}$ is related to the NSVZ beta function constraint.\footnote{In the case $N_c=1$ the rank-2 fields are absent and hence the factor $U(1)_a$ is absent. This comment applies to each theory we discuss in this paper, except for section \ref{sec:AD}.} For simplicity, throughout the paper, we simply write $SU(8) \times U(1)$ (or its analogs) and omit the $U(1)_R$ factor.

For the duality \ref{BASEDUAL}, the chiral ring generators map as
\be \ba{cccc}
 \left\{ \ba{c}tr( q_i a^r q_j) \\  tr( a^l ) \ea  \right\} &\Longleftrightarrow&   \left\{\ba{c} \mu_{ij; N_c-r-1} \\  tr( A^l ) \ea  \right\}
  \qquad \ba{l}  i<j=1,\ldots,8 \\ l=2,\ldots,N_c \ea
   \ea \ee
Generically, for small $r$ and $l$, these operators violate the unitarity bound $R>\frac{2}{3}$, so they are not in the chiral ring and should be flipped introducing appropriate gauge singlets $\b_{\CO}$ as prescribed in \cite{Benvenuti:2017lle}.

In the case of $N_c=1$, \ref{BASEDUAL} reduces to the $4d$ Intriligator-Pouliot duality for $SU(2)$ gauge theory (in the Intriligator-Pouliot duality the magnetic theory has $\binom{N_f}{2}$ gauge singlets). The $SU(2)$ gauge theories also enjoy Seiberg duality, where the magnetic theory has $\left(\frac{N_f}{2}\right)^2$ gauge singlets (we are counting real flavors here). In analogy, all the $Usp(2N_c)$ theories with antisymmetric and $8$ flavors satisfy other dualities, which we now describe.

Let us call \ref{BASEDUAL} \emph{Intriligator-Pouliot like} duality. The \emph{Seiberg like} duality splits the $8$ real flavors into $4$ "fundamentals" and $4$ "anti-fundamentals", and on the magnetic side only the "mesons" are flipped, so there are $4^2=16$ towers of singlets $m_{ij;r}$:
  \be \label{BASEDUALseiberg} \ba{ccc}
\ba{c}Usp(2N_c)\, \text{w/ antisymmetric} \, a \\
 \textrm{and $8$ flavors} \, q_i. \\
   \CW_{4d}=0 \ea 
    &\Longleftrightarrow& 
\ba{c}Usp(2N_c)\, \text{w/ antisymmetric} \, A, \\
 \textrm{ $4\!+\!\bar{4}$ flavors $Q_i,\Qt_i$ and singlets $m_{ij;r}$}. \\
   \CW_{4d}=\sum_{r=0}^{N_c-1}\sum_{i,j=1}^4 m_{i j; r}tr(\Qt_iA^r Q_j) \ea 
   \ea \ee
The UV global symmetry on the r.h.s. is only $SU(4)^2 \times U(1)^2$. The chiral ring generators map as
\be \ba{cccc}
 \left\{ \ba{c}tr( q_i a^r q_j) \\ tr( q_i a^r q_j) \\ tr( q_i a^r q_j) \\  tr( a^l ) \ea  \right\} &\Longleftrightarrow&   
 \left\{\ba{c} tr(Q_iA^r Q_j) \\ m_{ij; N_c-r-1} \\ tr(\Qt_iA^r \Qt_j) \\  tr( A^l ) \ea  \right\}
 \qquad \ba{l}  i<j=1,\ldots,4 \\  i=1,\ldots,4,\, j=5,\ldots,8 \\  i<j= 5,\ldots,8 \\ l=2,\ldots,N_c \\ \ea
   \ea \ee

Combining \ref{BASEDUAL} and \ref{BASEDUALseiberg}, one finds a third duality, which also splits the $8$ flavors into $4$ "fundamentals" and $4$ "anti-fundamentals", and in the magnetic side only the "baryons" are flipped, so there are $28-16=12$ towers of singlets $b_{i<j;r}, \bt_{i<j;r}$ ($i,j=1,\ldots,4$). Similarly to \cite{Dimofte:2012pd}, we call it \emph{ Csaki et al. like} duality:
  \be \label{BASEDUALcsaki} \ba{ccc}
\ba{c}Usp(2N_c)\, \text{w/ antisymmetric} \, a \\
 \textrm{and $8$ flavors} \, q_i. \\
   \CW_{4d}=0 \ea 
    &\Longleftrightarrow& 
\ba{c}Usp(2N_c)\, \text{w/ antisymmetric} \, A, \\
 \textrm{ $4\!+\!\bar{4}$ flavors $Q_i,\Qt_i$ and singlets $b_{ij;r}, \bt_{ij; r}$}. \\
   \CW_{4d}=\sum_{r=0}^{N_c-1}\sum_{i<j=1}^4 \left(b_{i j; r}tr(Q_iA^r Q_j) + \bt_{i j; r}tr(\Qt_iA^r \Qt_j)\right) \ea 
   \ea \ee
Also in this case the UV global symmetry on the r.h.s. is only $SU(4)^2 \times U(1)^2$. The chiral ring generators map as
\be \ba{cccc}
 \left\{ \ba{c}tr( q_i a^r q_j) \\ tr( q_i a^r q_j) \\ tr( q_i a^r q_j) \\  tr( a^l ) \ea  \right\} &\Longleftrightarrow&   
 \left\{\ba{c} b_{ij; N_c-r-1} \\  tr(\Qt_iA^r Q_j) \\ \bt_{ij; N_c-r-1} \\  tr( A^l ) \ea  \right\}
 \qquad \ba{l}  i<j=1,\ldots,4 \\  i=1,\ldots,4,\, j=5,\ldots,8 \\  i<j= 5,\ldots,8 \\ l=2,\ldots,N_c \\ \ea
   \ea \ee

Considering all the different ways to split the $8$ flavors into $4\!+\!\bar{4}$, it is actually possible to produce a total of $72$ different dual frames \cite{Rains:2003, Spiridonov:2008zr, Dimofte:2012pd}: $1$ frame with no singlets, $1$ \emph{Intriligator-Pouliout like} frame with $28N_c$ singlets, $\frac{1}{2}\binom{8}{4}=35$ \emph{Seiberg like} frames with $16 N_c$ singlets and $35$  \emph{Csaki et al. like} frames with $12 N_c$ singlets. This set of $72$ theories is parametrized by the coset $W(E_7)/W(SU(8))$, where $W(G)$ is the Weyl group of the Lie group $G$. 

\subsubsection*{The most fundamental duality}
Here we show that the  \emph{Csaki et al. like} duality \ref{BASEDUALcsaki} can be iterated to obtain the other two dualities \ref{BASEDUAL} and \ref{BASEDUALseiberg}. 

Let us consider the same \ref{BASEDUALcsaki} duality twice, but in one case relabeling $q_3,q_4 \rightarrow q_5,q_6$, in other words on the r.h.s. we have flipping fields 
\be \left( \ba{cccccccc} 
 \phantom{-} & \mu_{12;r} & \phantom{-}  & & \mu_{15;r} & \mu_{16;r} & & \\
& &  & & \mu_{25;r} & \mu_{26;r} &  & \\
& && \mu_{34;r} &   &  &\mu_{37;r} & \mu_{38;r} \\
& & & &  &  &\mu_{47;r} & \mu_{48;r} \\
& & & &  &\mu_{56;r}  && \\
& & & &  &  && \mu_{78;r}\\
& & & &  & && 
\ea \right) \ee
In the second case, in \ref{BASEDUALcsaki}, we relabel $q_3,q_4 \rightarrow q_7,q_8$, so on the r.h.s. we have flipping fields 
\be \left( \ba{cccccccc} 
\phantom{-} & \mu_{12;r} &\phantom{-} & &  & &\mu_{17;r} & \mu_{18;r} \\
& &  & &&  & \mu_{27;r} & \mu_{28;r}  \\
& & & \mu_{34;r} &\mu_{35;r} & \mu_{36;r} &  & \\
& & &  & \mu_{45;r} & \mu_{46;r}  && \\
& & & &  &\mu_{56;r}  && \\
& & & &  &  && \mu_{78;r}\\
& & & &  & && 
\ea \right) \ee
Composing these two dualities produces a new duality where the mesons in the $12, 34, 56, 78$ positions are flipped twice (which is equivalent to no flip, so these singlet fields disappear), and we are left with $12+12-2\cdot4=16$ towers of flipping fields, sitting in the $ij$ positions, with $i=1,\ldots,4$ and $j=5,\ldots,8$. This is precisely the \emph{Seiberg like} duality \ref{BASEDUALseiberg}. In other words we can write
%\be \left( \ba{cccccccc} 
%&\, &\, &\, &  \mu_{15;r} & \mu_{16;r} &\mu_{17;r} & \mu_{18;r} \\
%& &  & & \mu_{25;r} & \mu_{26;r} & \mu_{27;r} & \mu_{28;r}  \\
%& & && \mu_{35;r} & \mu_{36;r} &  \mu_{37;r} & \mu_{38;r} \\
%& &   &  & \mu_{45;r} & \mu_{46;r} & \mu_{47;r} & \mu_{48;r} \\
%& & & &  &  && \\
%& & & &  &  && \\
%& & & &  & && 
%\ea \right) \ee
\be \bigg(\text{\emph{ Csaki et al. like} duality \ref{BASEDUALcsaki}}\bigg)^2  = \text{\emph{Seiberg like} duality \ref{BASEDUALseiberg}} \ee

Composing the \emph{Seiberg like} duality with the \emph{ Csaki et al. like} duality produces the \emph{Intriligator-Pouliot like} duality, so we can also write:
\be \bigg(\text{\emph{ Csaki et al. like} duality \ref{BASEDUALcsaki}}\bigg)^3  =  \text{\emph{Intriligator-Pouliot like} duality \ref{BASEDUAL}} \ee

\subsection{$Usp(2N_c)$ with $6$ flavors, $\CW=0$}
All of the above is true also in $4d$, now we begin the study of truly $3d$ models.

Starting from the  \emph{Intriligator-Pouliot like} duality \ref{BASEDUAL} compactified to $3d$, and turning on real masses $+1$ for $q_7$ and $-1$ for $q_8$ we break the  global symmetry to 
\be SU(8) \times U(1) \rightarrow SU(6) \times U(1)_a \times U(1)_q \ee
where $U(1)_a$  acts on the antisymmetric field $a$ and  $U(1)_q$ acts on the $6$ massless flavors. The presence of two independent $U(1)$ axial symmetries is related to the fact that the linear monopole superpotential disappears on both sides. This implies that there is a Coulomb branch: the $N_c$ dressed monopoles $\M_{a^l}$ ($\M_{a^l}$ is the monopole dressed with $r$ powers of the antisymmetric field $a$, $l=0,\ldots,N_c-1$) are in the chiral ring on the electric side.

On the r.h.s. the flavors $Q_7, Q_8$ and the singlets $\mu_{i7;r}, \mu_{i8;r}$ ($r=1,..,6$) are massive, while the flavors $Q_j$ and the singlets $\mu_{78;r}$ and $\mu_{ij; r}$ ($i<j=1,\ldots,6$) stay massless. 

The tower of  singlets $\mu_{78; r}$ does not couple to dressed mesons $tr(Q_i \Phi^r Q_j$) anymore, but interactions with monopoles are now allowed, and new superpotential terms preserving the $SU(6) \times U(1)_a \times U(1)_q$ symmetry are generated:
\be \delta \CW_{quantum}=  \sum_{r=0}^{N_c-1} \M_{A^r} \mu_{78; N_c-1-r} \ee
so on the magnetic side, $\M_{A^r}$ (the monopole dressed with $r$ powers of the antisymmetric field $A$) is set to zero in the chiral ring by the $\CF$-terms of the singlets.  In other words the $\mu_{78;r}$ gauge singlets become flipping fields for the dressed monopoles. 

The new IR duality reads
  \be \label{usp60Wflavdual} \ba{ccc}
\ba{c}Usp(2N_c)\, \text{w/ antisymmetric} \, a \\
 \textrm{and $6$ flavors} \, q_i. \\
   \CW=0 \ea 
    &\Longleftrightarrow& 
\ba{c}Usp(2N_c)\, \text{w/ antisymmetric} \, A, \\
 \textrm{ $6$ flavors $Q_i$ and singlets $\mu_{ij;r}$}. \\
   \CW=\sum_{r=0}^{N_c-1} \M_{A^r} \mu_{78; N_c-1-r}+\\+\sum_{r=0}^{N_c-1}\sum_{i<j=1}^6 \mu_{ij; r}tr(Q_iA^r Q_j) \ea 
   \ea \ee
The case $N_c=1$ was discovered in \cite{Aharony:1997gp}.

The $\CZ_{S^3}$ integral identity associated with this duality is Theorem 5.6.14 of \cite{FOKKO}.

\subsubsection*{A different $Usp(2N_c)$ dual}
We can perform the same procedure of giving real masses $+1$ for $q_7$ and $-1$ for $q_8$ to the \emph{ Csaki et al. like}  duality \ref{BASEDUALcsaki} (that has $12$ towers of singlets on the r.h.s.). 

On the r.h.s. the UV global symmetry breaks as $SU(4)^2 \times U(1)^2 \rightarrow SU(4) \times SU(2) \times U(1)^3$. The quarks $\Qt_3, \Qt_4$ and the singlets $\bt_{i3;r},\bt_{i4;r}$ ($i=1,2$) become massive. The tower of singlets $\bt_{12; r}$ pairs with the tower of dressed monopoles $\M_{A^r}$ in a quantum generated superpotential.

The new IR duality reads
  \be \label{usp60Wflavdual2} \ba{ccc}
\ba{c}Usp(2N_c)\, \text{w/ antisymmetric} \, a \\
 \textrm{and $6$ flavors} \, q_i. \\
   \CW=0 \ea 
    &\Longleftrightarrow& 
\ba{c}Usp(2N_c)\, \text{w/ antisymmetric} \, A, \\
 \textrm{ $4\!+\!\bar{2}$ flavors $Q_i, \Qt_j$}. \\
   \CW=\sum_{r=0}^{N_c-1} \M_{A^r} \bt_{34; N_c-1-r}+\\+\sum_{r=0}^{N_c-1}\sum_{i<j=1}^4 b_{ij; r}tr(Q_iA^r Q_j)
   +\\+\sum_{r=0}^{N_c-1} \bt_{12; r}tr(\Qt_1 A^r \Qt_2) \ea 
   \ea \ee

The $\CZ_{S^3}$ integral identity associated with this duality is Theorem 5.6.11 of \cite{FOKKO}.

There is also a $Usp(2N_c) \leftrightarrow U(N_c)_{\CW=\M^+\!+\!\M^-}$ dual, which can be obtained turning on real masses in the self-duality for $Usp(2N_c)$ with $8$ flavors. We do not discuss this duality here, see section \ref{sec:axial1}.

\subsection{$U(N_c)$ with $4$ flavors, $\CW=\M^++\M^-$}
A dual model to $U(N)$ gauge theory with monopole superpotential $\M^+ + \M^-$ (without the adjoint field, but with an arbitrary number of flavors $N_f$) was proposed in \cite{Benini:2017dud} (see also \cite{Amariti:2018gdc}).\footnote{A brane construction of such gauge theories and dualities was found in \cite{Amariti:2017gsm}, using brane setups similar to the ones introduced in \cite{Amariti:2015yea, Amariti:2015mva,Amariti:2015xna}.}, and it was obtained turning on real masses in the $3d$ reduction of the Intriligator-Pouliot $4d$ duality for $Usp(2N)$ with $2N_f$ flavors. 

Here we start from the \emph{Intriligator-Pouliot like} duality  \ref{BASEDUAL}, that also has matter in a rank-$2$ representation of the gauge group, but we can  follow the same steps of \cite{Benini:2017dud}: we split the $8$ flavors of the $Usp(2N_c)$ gauge theory in two sets of $4$ flavors, and give positive real mass $+1$ to the first $4$ flavors and negative real mass $-1$ to the second $4$ flavors.

The real masses break the continuos global symmetry 
\be SU(8) \times U(1) \rightarrow SU(4) \times SU(4) \times U(1) \,,\ee 
where the $U(1)$ is a combination of the two 'axial' $U(1)$'s, $U(1)_q$ and $U(1)_\phi$, which act on the three set of fields $\{q, \qt, \phi\}$ with charges $\{1,1,0\}$ and $\{0,0,1\}$ respectively.% (Similar for $\{Q, \Qt, \Phi\}$ on the r.h.s.).

A non trivial part of the story is that we focus on a vacuum in which both the gauge groups are Higgsed as
\be Usp(2N_c) \rightarrow U(N_c)\,, \ee
in other words we are on the Coulomb branch of the original $Usp$ theories, away from the origin. On this vacuum, each $Usp(2N_c)$ flavor (that has $2N_c$ components) is split into two $U(N_c)$ chiral fundamentals, one fundamental is massive, one fundamental is massless. On the electric side, the first $4$ real $Usp(2N_c)$ flavors $q_i$ reduce to $4$ $U(N_c)$ fundamentals $q_i$ ($i=1,\ldots,4$), while the last $4$ real $Usp(2N_c)$ flavors $q_i$ ($i=5,\ldots,8$) reduce to $4$ $U(N_c)$ anti-fundamentals $\qt_j$, that we label with $j=1,\ldots,4$. Out of the $\binom{2N_c}{2}-1$ components of the antisymmetric field $a$, $N_c^2-1$ are massless, giving an adjoint of $U(N_c)$ $\phi$. So at low energies we are left with $U(N_c)$ with an adjoint $\phi$ and $(4,4)$ fundamentals $q_i, \qt_i$. 

On the magnetic side the story is similar, but there are also the gauge singlets. The singlets $\mu_{ij; r}$ are massless if $i=1,\ldots,4$ and $j=5,\ldots,8$, so they remain part of the magnetic theory at low energies. The $\mu_{ij; r}$ with $i<j=1,\ldots, 4$ or $i<j=5,\ldots,8$ are instead massive, and disappear.

Monopole terms in the superpotential are generated because of the Higgsing of the gauge group, through the Polyakov-Affleck-Harvey-Witten mechanism:
\be \delta \CW_{quantum}= \M^+ + \M^-\ee
The two linear monopole superpotential terms break the $U(1)_{top}$ topological symmetry of $U(N_c)$ and one of the two abelian axial symmetries, so they ensure that the global symmetries are correct. Moreover, monopole superpotential lifts completely the Coulomb branch: no monopole operators are in the chiral ring on either side of the duality.

Putting all these ingredients together, in the IR we get a $3d$ $U(N_c) \leftrightarrow U(N_c)$ duality with $16$ towers of gauge singlets $\mu_{ij;r}$:
  \be \ba{ccc}\label{U4flavdual}
\ba{c}U(N_c)\, \text{w/ adjoint} \, \phi \,
 \textrm{and} \, 4 \, \textrm{flavors}  \, q_i,\qt_j \\
   \CW=\M^+ + \M^- \ea 
    &\Longleftrightarrow& 
\ba{c}U(N_c)\, \text{w/ adjoint} \, \Phi \, 
 \textrm{and} \, 4 \, \textrm{flavors} \, Q_i, \Qt_j \\
   \CW=\M^++\M^-+ \qquad\qquad  \\ \quad +\sum_{r=0}^{N_c-1}\sum_{i,j=1}^4 \mu_{ij; r}tr(Q_i \Phi^{r}\Qt_j) \ea 
   \ea \ee
 where $i,j=1,\ldots,4$.  
The chiral ring generators map as
\be \ba{ccc}
 \left\{\ba{c}tr( q_i \phi^r \qt_j) \\  tr( \phi^j ) \ea  \right\}&\Longleftrightarrow&  \left\{ \ba{c} \mu_{ij; N_c-r-1} \\  tr( \Phi^j ) \ea  \right\}
   \ea \ee
The $N_c=1$ case was discussed in \cite{Dimofte:2012pd}. 

\subsubsection*{A different $U(N_c)$ dual}
We can perform the same procedure of giving real masses starting from the \emph{ Csaki et al. like} duality  \ref{BASEDUALcsaki}.\footnote{We can also consider the same real mass deformation on all the theories in the duality web of $72$ models. The discussion of \cite{Dimofte:2012pd} for the $N_c=1$ case generalizes in a straightforward way. The $72$ models of $Usp(2N_c)\, \textrm{w}/ 2N_f\!=\!8$ reduce to 32 theories $Usp(2N_c) \, \textrm{w/}\, 6$ fundamentals (labelled by elements of the coset $W(SO(12)) / W(SU(6))$) plus 40 theories $U(N_c)\,  \textrm{w/}\,4$ flavors (labelled by elements of the coset $W(SO(12)) / W(SU(4) \times SU(4))$). The $E_7$ action is broken to a $D_6=SO(12)$ action. The $40$ $U(N_c)$ theories are actually collapsed to $20$: $1$ theory with no singlets, $1$ theory with $16 N_c$ singlets, $18$ theories with $8 N_c$ singlets ($18=\frac{1}{2}\binom{4}{2}^2$ is the number of ways one can split the $(4,4)$ flavors into $(2+\tilde{2},2+\tilde{2})$) flipping the $8$ "baryons".} We give real mass $+1$ to the flavors $1,2,\tilde{1},\tilde{2}$ and real mass $-1$ to the flavors $3,4,\tilde{3},\tilde{4}$. Out of the $12$ towers of "baryon"-singlets $b_{i<j;r}, \bt_{i<j;r}$, $4$ towers ($b_{12;r}, b_{34;r}, \bt_{12;r}, \bt_{34;r}$) become massive, the other $8$ towers stay massless. Also in this case a monopole superpotential $\M^++\M^-$ is generated. Relabelling the r.h.s. fields, in the IR we get the following $U(N_c) \leftrightarrow U(N_c)$ duality with $4+4$ towers of gauge singlets $b_{ij;r}$ and $\bt_{ij;r}$:
  \be \ba{ccc}\label{U4flavdual2}
\ba{c}U(N_c)\, \text{w/ adjoint $\phi$} \\ \text{and $(4,4)$ flavors}  \, q_i,\qt_j \\
   \CW=\M^+ + \M^- \ea 
    &\Longleftrightarrow& 
\ba{c}U(N_c)\, \text{w/ adjoint $\Phi$ and}\\\text{$(2\!+\!\bar{2},2\!+\!\bar{2})$ flavors} \, Q_i, \Qt_j \\
   \CW=\M^++\M^-+ \qquad\qquad\qquad  \\ \quad +\sum_{r=0}^{N_c-1}\sum_{i=1}^2\sum_{j=3}^4 b_{ij; r}tr(Q_i \Phi^{r}\Qt_j)+ \\
   \quad +\sum_{r=0}^{N_c-1} \sum_{i=3}^4\sum_{j=1}^2 \bt_{ij; r}tr(Q_i \Phi^{r}\Qt_j) \,\,\,  \ea 
   \ea \ee
On the r.h.s. UV global symmetry is only $SU(2)^4 \times U(1)^3$. The chiral ring generators map as
\be \ba{cccc}\label{mapmirrorU4F}
 \left\{\ba{c}
 tr( \qt_i \phi^r q_j) \,,\,\, tr( \qt_{i} \phi^r q_{j+2}) \\  
 tr( \qt_{i+2} \phi^r q_j) \,,\,\, tr( \qt_{i+2} \phi^r q_{j+2})  
  \\ tr( \phi^l ) \ea  \right\}
 &\Longleftrightarrow&
   \left\{ \ba{c} tr( \Qt_i \Phi^r Q_j) \,,\quad   b_{ij; N_c-r-1}\\ 
                    \bt_{ij; N_c-r-1} \,,\,\,  tr( \Qt_{i+2} \Phi^r Q_{j+2} )  \\
                      tr( \Phi^l )  \ea  \right\} &
   \ba{c} i,j=1,2 \\ i,j=1,2 \\ l=2,\ldots,N_c\ea
   \ea \ee
 The $N_c=1$ case was discussed in \cite{Dimofte:2012pd}. The $U(1)_{\CW=\M^+\!+\!\M^-}$ theory with $4$ flavors and $8$ flipping fields appears as a boundary condition for $\CN=2$ rank-$1$ class-S SCFT's \cite{Bawane:2017gjf}.

\subsection{$U(N_c)$ with $3$ flavors, $\CW=\M^+$}
Now we start from \ref{U4flavdual} and turn on real and opposite masses for $q_4$ and $\qt_4$, following the same strategy of section 8 of \cite{Benini:2017dud}. On the r.h.s. the flavors $Q_4$ and $\Qt_4$, and the singlets $\mu_{i4;r}$ and $\mu_{4i;r}$ (for $i=1,2,3$) becomes massive. The singlets $\mu_{ij;r}$ and $\mu_{44;r}$ stay massless. We rename $\mu_{44;r} \rightarrow \mu_{-;r}$.

The $SU(4)^2 \times U(1)$ global symmetry breaks to $SU(3)^2 \times U(1) \times U(1)$, so the process removes one linear monopole superpotential term (if both monopole terms $\M^+$ and $\M^-$ are present, the new $U(1)$ symmetry would be broken). 
 New superpotential terms preserving all the magnetic side symmetries are generated:
\be \delta \CW_{quantum}= \sum_{r=0}^{N_c-1} \mu_{-; r}\M^+_{\Phi^r} \ee
We denote the monopole operators dressed by $k$ powers of the adjoint field $\phi$ as $\M_{\phi^k}$. For $U(N_c)$ gauge group, all the dressed monopoles can be written as an algebraic combination of the $2N_c$ monopoles $\M^\pm_{\phi^k}$, $k=0,\ldots,N_c-1$ \cite{Cremonesi:2013lqa}.

The duality reduces in the IR to
  \be \ba{ccc}\label{U3flavdual}
\ba{c}U(N_c)\, \text{w/ adjoint} \, \phi \\
 \textrm{and $3$ flavors} \,q_i,\qt_i \\
   \CW=\M^+  \ea 
    &\Longleftrightarrow& 
\ba{c}U(N_c)\, \text{w/ adjoint} \, \Phi \\
 \textrm{and $3$ flavors}\, Q_i, \Qt_i \\
   \CW=\M^-+ \sum_{r=0}^{N_c-1} \mu_{-; r}\M^+_{\Phi^r} +\\ \quad +\sum_{r=0}^{N_c-1}\sum_{i,j=1}^3 \mu_{ij; r}tr(Q_i \Phi^r\Qt_j) \ea 
   \ea \ee
The surviving gauge singlets on the r.h.s.  now flip both the mesons and the dressed monopoles $\M^+_{\Phi^r}$, while the tower of dressed monopoles $\M^-_{\Phi^r}$ is not in the chiral ring because of the superpotential term $\M^-$. 

The chiral ring generators map as
\be \ba{ccc}\label{U3flavmap}
 \left\{\ba{c}tr( q_i \phi^r \qt_j)  \\ \M^-_{\phi^r} \\  tr( \phi^l )\ea  \right\}&\Longleftrightarrow&  
 \left\{ \ba{c} \mu_{ij; N_c-r-1} \\ \mu_{-;N_c-1-r} \\  tr( \Phi^l ) \ea  \right\}
   \ea \ee
   
In  the case $N_c=1$ this duality appeared in \cite{Benini:2017dud}, as a special case of a duality valid for $U(N_c)$ without adjoint, but with an arbitrary number of flavors $N_f$ \cite{Benini:2017dud}.

Let us mention that the duality \ref{U3flavdual} can also be reached starting from the $Usp \leftrightarrow Usp$ duality \ref{usp60Wflavdual}, turning on real masses $(+,+,+,-,-,-)$ for the $6$ flavors on the electric side. On both sides the gauge groups Higgs down to $U(N_c)$, and the massless magnetic singlets are $\mu_{78; r}$ and $\mu_{ij; r}$ with $i=1,2,3$, $j=1,2,3$.

\subsubsection*{A different $U(N_c)$ dual}
We can also  start from the duality \ref{U4flavdual2}, and we get a duality different from \ref{U3flavdual}. We turn on real and opposite masses for $q_4$ and $\qt_4$, so that $b_{14; r}, b_{24; r}, b_{41; r}, b_{42; r}$ become massive, while $b_{13; r}, b_{23; r}, \bt_{31; r}, \bt_{32; r}$ stay massless and keep flipping the "baryons". This time no quantum superpotential is generated, and the duality reads
  \be \ba{ccc}\label{U3flavdual2}
\ba{c}U(N_c)\, \text{w/ adjoint} \, \phi \\
 \textrm{and $(3,3)$ flavors} \,q_i,\qt_i \\
   \CW=\M^+  \ea 
    &\Longleftrightarrow& 
\ba{c}U(N_c)\, \text{w/ adjoint} \, \Phi \\
 \textrm{and $(2\!+\!\bar{1},2\!+\!\bar{1})$ flavors} \, Q_i, \Qt_i \\
   \CW=\M^- + \sum_{r=0}^{N_c-1}\sum_{i=1}^2 b_{i3; r} tr(Q_i \Phi^r\Qt_3)+\\+ \sum_{r=0}^{N_c-1}\sum_{i=1}^2 \bt_{3i;r} tr(Q_3 \Phi^r\Qt_i) \ea 
   \ea \ee
On the r.h.s. UV global symmetry is only $SU(2)^2 \times U(1)^4$. The chiral ring generators map as
\be \ba{ccc}
 \left\{\ba{c}
 tr( q_1 \phi^r \qt_1) \,,\,\, tr( q_1 \phi^r \qt_2) \,,\,\, tr( q_1 \phi^r \qt_3) \\  
 tr( q_2 \phi^r \qt_1) \,,\,\, tr( q_2 \phi^r \qt_2) \,,\,\, tr( q_2 \phi^r \qt_3)  \\ 
  tr( q_3 \phi^r \qt_1) \,,\,\, tr( q_3 \phi^r \qt_2) \,,\,\, tr( q_3 \phi^r \qt_3) \\
   \M^-_{\phi^r} \\ tr( \phi^j ) \ea  \right\}
 &\Longleftrightarrow&
   \left\{ \ba{c} tr( Q_1 \Phi^r \Qt_1) \,,\,\,  tr( Q_1 \Phi^r \Qt_2) \,,\,\, b_{13; N_c-r-1}\\ 
   tr( Q_2 \Phi^r \Qt_1) \,,\,\,  tr( Q_2 \Phi^r \Qt_2) \,,\,\, b_{23; N_c-r-1} \\
  \bt_{31; N_c-r-1} \,\, ,\,\,\,\, \bt_{32; N_c-r-1} \,,\quad \M^+_{\Phi^r} \\ tr( Q_3 \Phi^r \Qt_3) \\  tr( \Phi^j )  \ea  \right\}
   \ea \ee
As far as we know, duality \ref{U3flavdual2} is new also in the case $N_c=1$. 

Let us emphasize that the above mapping and the mapping \ref{mapmirrorU4F} resemble the mapping of mirror symmetries \cite{Intriligator:1996ex}: in mirror symmetries monopoles map to mesons, while mesons map to either singlets, mesons or monopoles. We will indeed see in section \ref{sec:UN2Fmirror} that \ref{U3flavdual2} can be deformed to a mirror symmetry like duality, with a  brane setup such that the gauge theory duality corresponds to S-duality in the brane setup. 
   
\subsection{$U(N_c)$ with $2$ flavors, $\CW=0$}\label{sec:U2flavoW}
We start from \ref{U3flavdual}, in order to remove the last monopole superpotential, we turn on real and opposite masses for $q_3$ and $\qt_3$. 

The continuos global symmetry breaks as
\be SU(3)^2 \times U(1)^2 \rightarrow SU(2)^2 \times U(1)_{top} \times U(1)_q \times U(1)_\phi \ee
As before, the superpotential terms linear in the monopoles disappear on both sides, while a new tower of monopole-flipping superpotential terms is generated on the r.h.s. 

On the r.h.s. of \ref{U3flavdual}, the flavors $Q_3$ and $\Qt_3$, and the singlets $\mu_{i3;r}$ and $\mu_{3i;r}$ (for $i=1,2$) becomes massive. The singlets $\mu_{ij; r}, \mu_{33; r}, \mu_{-; r}$ ($i,j=1,2$) stay massless. 

Renaming $\mu_{33;r} \rightarrow \mu_{-;r}$, at low energy we get a \emph{Aharony like} duality:
  \be \ba{ccc}\label{U2flavdual}
\ba{c}U(N_c)\, \text{w/ adjoint} \, \phi \\
 \textrm{and $2$ flavors} \, q_i,\qt_i \\
   \CW=0  \ea 
    &\Longleftrightarrow& 
\ba{c}U(N_c)\, \text{w/ adjoint} \, \Phi \\
 \textrm{and $2$ flavors}\, Q_i,\Qt_i \\
   \CW= \sum_{r=0}^{N_c-1} \mu_{\mp; r}\M^{\pm}_{\Phi^r} +\\ \quad +\sum_{r=0}^{N_c-1}\sum_{i,j=1}^2 \mu_{ij; r}tr(Q_i \Phi^r \Qt_j) \ea 
   \ea \ee
If $N_c=1$ the duality reduces to the Aharony duality for $U(1)$ with $2$ flavors \cite{Aharony:1997gp}. The chiral ring generators map as
\be \ba{ccc}\label{U2flavmap}
 \left\{\ba{c}tr( q_i \phi^r \qt_j) \\  tr( \phi^j ) \\ \M^\pm_{\phi^i} \ea  \right\}&\Longleftrightarrow&  \left\{ \ba{c} \mu_{ij; N_c-r-1} \\  tr( \Phi^j ) \\ \mu_{\pm;N_c-i-1} \ea  \right\}
   \ea \ee
Let us display the charges of the relevant fields under the global symmetry $SU(2)^2 \times U(1)_{top} \times U(1)_q \times U(1)_{\phi}$ as
\be \ba{c||c|c|c|c|c||c}
                &  SU(2)  \otimes SU(2) &U(1)_{q} & U(1)_{\phi} & U(1)_{top} & U(1)_R &\\\hline
 q_i,\qt_i   & {\bf (2,1)} ,  {\bf (1,2)} &  1  & 0 &  0                 &   r_q     &\\  
\phi          &   ({\bf 1},{\bf 1})           &    0     & 1 &0    &     \rf  &\\   
\M^\pm    & ({\bf 1},{\bf 1})             &  -2 &  0 & \pm 1    &2-2r_q-(N_c-1)\rf &\\   \hline
               & {\bf (2,1)} ,  {\bf (1,2)}   & -1 & 0  &  0          &      1-r_q-\frac{(N_c-1) \rf }{2} & Q_i,\Qt_i \\   
                    &  ({\bf 1},{\bf 1})        &0 &  1& 0          &          \rf  & \Phi \\   
                     & ({\bf 1},{\bf 1})        &2&  0 & \mp 1   &        2r_q & \M^{\pm} \\   
                    &  ({\bf 2},{\bf 2})        &2&   0 &0                     &  2r_q+(N_c-1-r)\rf & \mu_{ij;r} \\   
                      &  ({\bf 1},{\bf 1})      &-2&   0 &\pm 1             & 2-2r_q - r\cdot \rf & \mu_{\pm;r} \\   
\ea \ee
The Abelian global symmetries $U(1)_q$ and $U(1)_\phi$ mix with the R-symmetry, accordingly $r_q$ and $r_\phi$ are the two independent variables appearing in $\CZ$-extremization, which can be used to determine the superconformal R-charges.

\subsubsection{A \emph{mirror like} duality and its brane setup}\label{sec:UN2Fmirror}
We can also find a different dual for $U(N_c)$ with adjoint and $(2,2)$ flavors, starting from \ref{U4flavdual2}, giving real mass $+1$ to $q_2$ and real mass $-1$ to $\qt_2$. $Q_2,\Qt_2$ and the singlets $b_{23;r}, \bt_{32;r}$ become massive, the other $2$ flavors and the singlets $b_{13;r}, \bt_{31;r}$ stay massless. 

Renaming $q_3, \qt_3 \rightarrow q_2, \qt_2$ and $\{Q_1, \Qt_1, Q_3, \Qt_3, b_{13;r}, \bt_{31;r} \} \rightarrow \{ Q_L, \Qt_R, Q_R, \Qt_L, \mu_{L; r}, \mu_{R; r} \}$ (so that the singlets flip the diagonal instead of the off-diagonal mesons), the low energy duality reads:
  \be \ba{ccc}\label{U2flavdual2}
\ba{c}U(N_c)\, \text{w/ adjoint} \, \phi \\
 \textrm{and $(2,2)$ flavors} \, q_i, \qt_i \\
   \CW=0  \ea 
    &\Longleftrightarrow& 
\ba{c}U(N_c)\, \text{w/ adjoint} \, \Phi \\
 \textrm{and $(1\!+\!\bar{1},1\!+\!\bar{1})$ flavors}\, Q_{L,R}, \Qt_{L,R} \\
   \CW\!=\! \sum_{r=0}^{N_c-1} \sum_{i=L,R} b_{i; r} tr(Q_i \Phi^r \Qt_i)  \ea 
   \ea \ee
 with mapping  
\be \ba{ccc}
 \left\{\ba{c}  tr(\qt_1 \phi^r q_1) ,  tr(\qt_1 \phi^r q_2)\\
 tr(\qt_1 \phi^r q_2) ,  tr(\qt_2 \phi^r q_2) \\ \M^+_{\phi^r} , \M^-_{\phi^r} \\ tr(\phi^r) \ea  \right\}
       &\Longleftrightarrow&  
  \left\{\ba{c}  b_{L;N_c-1-r}  \, , \,\,\M^{+}_{\Phi^r} \\ 
   \M^-_{\Phi^r} \, , \,\,   b_{R;N_c-1-r}\\
     tr(\qt_L  \Phi^r q_R) , tr(\qt_R \Phi^r \qt_L)  \\ tr(\Phi^r) \ea  \right\}
     \ea \ee

If $N_c=1$ the duality reduces to the $\CN\!=\!2$ mirror symmetry duality for $U(1)$ with $2$ flavors of \cite{Aharony:1997bx}.

\paragraph{Brane interpretation} The duality \ref{U2flavdual2} admits a  brane interpretation in Type IIB superstring \`a la Hanany-Witten \cite{Hanany:1996ie}:
%\begin{figure}
\be\label{PIC:Nf=2duality}
\begin{tikzpicture}
\draw [ultra thick, blue] (2,2.5) -- (2,-2.5);\node at (2.35,2.5) {NS};
\draw [dashed, ultra thick, gray] (0.3,2.5) -- (0.3,-2.5);\node at (0.65,2.5) {D5'};
\draw [dashed, ultra thick, gray] (-0.9,2.5) -- (-0.9,-2.5);\node at (-0.55,2.5) {D5'};
\draw [ultra thick, blue] (-2.5,2.5) -- (-2.5,-2.5);\node at (-2.15,2.5) {NS};
\node at (-1.7, 1.3){$N_c$};
\node at (-0.3, 1.3){$N_c$};
\node at (1.1, 1.3){$N_c$};
\draw [thick, red] (-2.5,1) -- (-0.9,1);
\draw [thick, red] (-2.5,0.7) -- (-0.9,0.7);
\draw [thick, red] (-2.5,0.4) -- (-0.9,0.4);
\draw [thick, red] (-2.5,-1) -- (-0.9,-1);
\draw [thick, dashed, red] (-2.5,-0.3) -- (2,-0.3);
\draw [thick, red] (2,0.8) -- (0.3,0.8);
\draw [thick, red] (2,0.5) -- (0.3,0.5);
\draw [thick, red] (2,0.3) -- (0.3,0.3);
\draw [thick, red] (2,-0.8) -- (0.3,-0.8);
\draw [thick, red] (-0.9,0.9) -- (0.3,0.9);
\draw [thick, red] (-0.9,0.55) -- (0.3,0.55);
\draw [thick, red] (-0.9,0.1) -- (0.3,0.1);
\draw [thick, red] (-0.9,-1.2) -- (0.3,-1.2);
\node at (4,0) {$\qquad\qquad\Longleftrightarrow\qquad\qquad$};
\node at (4,0.5) {S-duality};
\end{tikzpicture} 
%\begin{tikzpicture}\node at (0,2.5) {$\qquad\Longleftrightarrow\qquad$};\end{tikzpicture}
\begin{tikzpicture}
\draw [dashed, ultra thick, gray] (2,2.5) -- (2,-2.5);\node at (2.35,2.5) {D5'};
\draw [ultra thick, blue] (0.3,2.5) -- (0.3,-2.5);\node at (0.65,2.5) {NS};
\draw [ultra thick, blue] (-0.9,2.5) -- (-0.9,-2.5);\node at (-0.55,2.5) {NS};
\draw [dashed, ultra thick, gray] (-2.5,2.5) -- (-2.5,-2.5);\node at (-2.15,2.5) {D5'};
\node at (-1.7, 1.3){$N_c$};
\node at (-0.3, 1.3){$N_c$};
\node at (1.1, 1.3){$N_c$};
\draw [thick, red] (-2.5,1) -- (-0.9,1);
\draw [thick, red] (-2.5,0.7) -- (-0.9,0.7);
\draw [thick, red] (-2.5,0.4) -- (-0.9,0.4);
\draw [thick, red] (-2.5,-1) -- (-0.9,-1);
\draw [thick, dashed, red] (-2.5,-0.3) -- (2,-0.3);
\draw [thick, red] (2,0.8) -- (0.3,0.8);
\draw [thick, red] (2,0.5) -- (0.3,0.5);
\draw [thick, red] (2,0.3) -- (0.3,0.3);
\draw [thick, red] (2,-0.8) -- (0.3,-0.8);
\draw [thick, red] (-0.9,0.9) -- (0.3,0.9);
\draw [thick, red] (-0.9,0.55) -- (0.3,0.55);
\draw [thick, red] (-0.9,0.1) -- (0.3,0.1);
\draw [thick, red] (-0.9,-1.2) -- (0.3,-1.2);
\end{tikzpicture}  \ee%\end{figure}   
The $NS5$ branes stretch along $012345$, the $N_c$ $D3$ branes stretch along $0126$ and the $D5'$ branes along $012347$. The $N_c$ D3 branes give rise to a $U(N_c)$ gauge theory with a massless adjoint $\phi$ on the l.h.s. and $\Phi$ on the r.h.s. (more precisely $\phi$ and $\Phi$ have $N_c^2$ degrees of freedom so there is also a gauge singlet, which however decouples).

On the l.h.s. the central $D5'$'s provide the $2$ flavors $q_i,\qt_i$ to the $U(N_c)$ gauge theory ($D3-D5$ strings), which do not couple to $\phi$.

On the r.h.s. the $D5'$'s are outside the $NS$ branes, each $D5$ still provides a flavor, but on the r.h.s. there are $N_c+N_c$ complex degrees of freedom associated the vertical motion of the external $D3$ segments. Such d.o.f. are represented by $2$ towers of $N_c$ gauge singlets $b_{L; r}, b_{R; r}$, flipping the dressed mesons $tr(Q_L \Phi^r \Qt_L)$, $tr(Q_R \Phi^r \Qt_R)$.

The two brane setups in \ref{PIC:Nf=2duality} are related by Type IIB S-duality, so we can interpret the duality \ref{U2flavdual2} as a mirror symmetry, with $\CN\!=\!2$ supersymmetry. This suggests a natural generalization to a mirror duality for $\CN\!=\!2$ adjoint-$U(N_c)$ with $N_f$ flavors and $\CW=0$. We will discuss this mirror symmetry in section \ref{sec:mirror}.

It would be interesting to discover a brane interpretation of the dualities discussed in the rest of this section. Dualities like \ref{U4flavdual}, \ref{U3flavdual} and \ref{U2flavdual} involve monopoles in the superpotentials, so the relevant setups might be similar to the ones constructed in \cite{Amariti:2015yea, Amariti:2015mva,Amariti:2017gsm,Amariti:2015xna}.

%%%%%%%%%%%%%%%%%%%%%%%%%%%%%%%%%%%%%%%%%%%%%%%%%%%%%%%%%%%%%%
%%%%%%%%%%%%%%%%%%%%%%%%%%%%%%%%%%%%%%%%%%%%%%%%%%%%%%%%%%%%%%
%%%%%%%%%%%%%%%%%%%%%%%%%%%%%%%%%%%%%%%%%%%%%%%%%%%%%%%%%%%%%%
%%%%%%%%%%%%%%%%%%%%%%%%%%%%%%%%%%%%%%%%%%%%%%%%%%%%%%%%%%%%%%
\section{Turning on a complex mass: cubic Wess-Zumino duals}\label{sec:confining}
In this section we start from the dualities of section \ref{sec:selfduals} and turn on a complex mass for the electric theories (i.e. the side with no gauge singlet fields).  We consider giving mass to a minimal number of flavors, that is $2$ fundamentals for $Usp(2N_c)$ and $1$ fundamental and $1$ anti-fundamentals for $U(N_c)$.

 The dualities we start from have the following property: 
  \begin{quote} \emph{turning on a complex mass to a minimal number of flavors on the electric side of an \emph{Intriligator-Pouliot like} duality, the magnetic gauge group is completely broken.} \end{quote}
 This happens because on the magnetic side a superpotential term linear in a singlet field is turned on, and in order to satisfy the F-terms, the $2$-index field (antisymmetric for $Usp(2N_c)$ and adjoint for $U(N_c)$)  takes a maximal nilpotent vev. So we end up with a duality between a $Usp(2N_c)$ or $U(N_c)$ gauge theory and a theory with no gauge interactions (for this reason such dualities are usually called \emph{confining}). The gauge theories are:
 \be
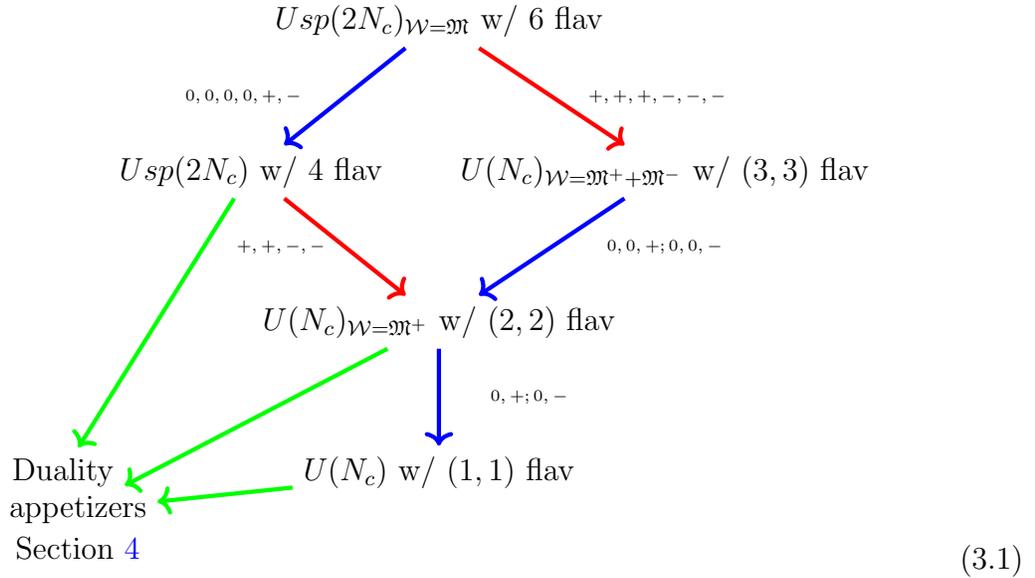
\label{WZscheme}
\begin{tikzpicture}
\node (a) at (0,2) {$Usp(2N_c)_{\CW=\M}$ w/ $6$ flav};
\node (b) at (-2.5,0) {$Usp(2N_c)$ w/ $4$ flav};
\node (c) at (3,0) {$U(N_c)_{\CW=\M^++\M^-}$ w/ $(3,3)$ flav};
\node (d) at (0,-2) {$U(N_c)_{\CW=\M^+}$ w/ $(2,2)$ flav};
\node (e) at (0,-4) {$U(N_c)$ w/ $(1,1)$ flav};
\draw [->,ultra thick, blue] (a) -- (b);\node at (2.9,1) {\tiny{$+,+,+,-,-,-$}};
\draw [->, ultra thick, red] (a) -- (c);\node at (-2.6,1) {\tiny{$0,0,0,0,+,-$}};
\draw [->,ultra thick, red] (b) -- (d);\node at (3,-1) {\tiny{$0,0,+;0,0,-$}};
\draw [->, ultra thick, blue] (c) -- (d);\node at (-2.1,-1) {\tiny{$+,+,-,-$}};
\draw [->,ultra thick, blue] (d) -- (e);\node at (1.2,-3) {\tiny{$0,+;0,-$}};
\node (f) at (-5,-4) {Duality};
\node (g) at (-4.8,-4.5) {appetizers};
\node at (-4.8,-5.0) {Section \ref{sec:appetizers}};
\draw [->,ultra thick, green] (b) -- (f);
\draw [->,ultra thick, green] (d) -- (g);
\draw [->,ultra thick, green] (e) -- (g);
\end{tikzpicture}\ee

A non trivial superpotential is generated for the low energy Wess-Zumino model, which we find as the most general interaction preserving all the symmetries. In each case the superpotential is cubic. This implies, in particular, that all the Wess-Zumino models are relevant deformations of a free theory and flow to a non-trivial SCFT in the IR.

For each model, the duality for the case $N_c=1$ already appeared in the literature.

We first discuss at length the case of $U(N_c)$ with $3$ flavors, $\CW=\M^++\M^-$, in section \ref{sec:Nf=3WZ}. The other cases are very similar and we skip many details. Let us also mention that each theory in \ref{WZscheme} (except for the top one, $Usp(2N_c)$ with $6$ flavors) can be obtained either by a complex mass deformation of the corresponding theory in \ref{selfdualscheme}, or by  a real mass deformation of a "higher" theory in \ref{WZscheme} (using the real masses indicated). This is consistent with the superpotential being always cubic.

\subsection{$U(N_c)$ with $3$ flavors, $\CW=\M^++\M^-$}\label{sec:Nf=3WZ}
If we start from the duality for $U(N_c)$ with $4$ flavors and $\CW=\M^++\M^-$, eq. \ref{U4flavdual}, and turn on a complex mass 
\be \delta\CW= tr(q_4\qt_4) \qquad \Longleftrightarrow \qquad \delta\CW=  \mu_{44;N_c-1}\,,\ee 
breaking the $SU(4)^2 \times U(1)$ global symmetry to $SU(3)^2 \times U(1)_{axial}$. 

On the l.h.s. the theory simply loses one flavor. 

On the r.h.s. the $\CF$-terms equations of $ \mu_{44;N_c-1}$ read $tr(Q_4\Phi^{N_c-1}\Qt_4)=-1$, so they imply that $Q_4$, $\Qt_4$ and $\Phi$ take a vev such that $tr(Q_4\Phi^{N_c-1}\Qt_4)$ is non-zero, while the other dressed mesons $tr(Q_4\Phi^{l}\Qt_4)$, $l=0,\ldots,N_c-2$, stay at zero vev. These $\CF$-terms are satisfied by a vacuum proportional to
\bea \label{VEV} Q_{1,2,3} \! & =& 0 \nn \\
 \Qt_{1,2,3} \!& = &0 \nn \\
\Qt_4 \, &=& (1, 0,\ldots,0) \\ 
\Phi \, &= & - \textrm{Jordan}_{N_c \times N_c } \nn \\
 Q_4 \, &= & (0,\ldots,0,1) \nn
\eea
The vev of $\Phi$ is a \emph{maximal} $N_c \times N_c$ Jordan block. Similar vev's were discussed in \cite{Kutasov:1995np,Kutasov:1995ss}. The vev's \ref{VEV} are such that $tr(\Qt_4\Phi^{r}Q_4)=\delta_{r, N_c-1}$, while the other dressed mesons $tr(\Qt_i\Phi^{r}Q_j)$ are zero.

On the vacuum \ref{VEV} the magnetic gauge symmetry breaks completely $U(N_c) \rightarrow U(0)$. The $6$ towers of singlets $\mu_{i4; r}, \mu_{4i, r}$ become massive. The low energy theory is a Wess-Zumino model with massless fields $\mu_{ij; r}$  ($i,j=1,2,3$, $r=0,\ldots,N_c-1$) and $\mu_{44; r}$ ($r=0,\ldots,N_c-2$). 

We now want to determine the precise superpotential of the Wess-Zumino model. If $N_c=1$ the fields $\mu_{44; r}$ are absent, the duality was studied in \cite{Benvenuti:2016wet} and the Wess-Zumino superpotential was found to be a determinant: 
\be \label{WwzU13F} N_c=1\,: \qquad \CW_{WZ}= det_{3\times 3}\left(\mu_{ij;0}\right) =  \sum_{i,j,k=1}^3 \varepsilon_{ijk} \,\, \mu_{1i; 0} \, \mu_{2j; 0} \, \mu_{3k; 0}\ee
preserving $SU(3) \times SU(3)$ invariance. The determinant is the only $SU(3)_L \times SU(3)_R$ invariant that can be made out of a $3 \times 3$ matrix.

\paragraph{Decoupling of the tower $\mu_{44; r}$.}
In order to figure out if for $N_c>1$ the tower of $N_c-1$ fields $\mu_{44;r}$ enter the superpotential of the WZ model, we performed $\CZ$-extremization on the theory adjoint-$U(N_c)$ with $3$ flavors and $\CW=\M^++\M^-$ (which is done on just on variable). If $N_c=2$, the result is $r_q= r_{\qt} = 0.2675$ and $r_{\phi}=0.1975$. This implies that $R[tr(\phi^2)]<\frac12$, so $tr(\phi^2)$ is a free field in the infrared. We are then led to apply the prescription of \cite{Benvenuti:2017lle}: introduce $N_c-1$ gauge singlets $\b_i$ that "flip" the fields $tr(\phi^j)$. On the dual side, the $N_c-1$ singlets $\mu_{44; r}$ become massive and disappear from the Wess-Zumino model.  

In \cite{Benvenuti:2017lle,Benvenuti:2017kud} (which studied a $4d$ duality that reduced to $3d$ becomes equivalent to the one we discuss in section \ref{sec:Nf=1}), the $\b_j$ fields cannot take a vev for quantum reasons: if such a field takes a vev, an ADS-like superpotential is generated, and there is no vacuum. This supports the conjecture that the $\b_j$'s are not in the quantum chiral ring. (Logically there is the possibility that the $\b_j$'s are nilpotent elements of the chiral ring, so they do not give rise to flat directions but are non-trivial chiral ring operators. In the analogous cases discussed in \cite{Benvenuti:2017lle,Benvenuti:2017kud,Benvenuti:2017bpg} this possibility was discarded showing that the $\b_j$'s are descendants under an emergent IR additional supersymmetry.)

We can also argue that the $N_c-1$ fields $\mu_{44; r}$ are free in the IR on the dual side: on the Wess-Zumino model (as we will see in the next paragraph), the $9N_c$ $\mu_{ij; r}$ fields enter cubically the superpotential. In order to preserve all the $SU(3)^2 \times U(1)_{axial}$ symmetries, the $\mu_{44; r}$ fields should enter the superpotential via terms of higher degree, that (in the absence of gauge interactions) are however always irrelevant, so the $\mu_{44; r}$'s would decouple from the rest of the SCFT.

\paragraph{The most general superpotential.}
We then need to generalize \ref{WwzU13F} and find the superpotential for $N_c>1$, as a function of the $\mu_{ij; r}$. Cubic couplings similar to a $3 \times 3$ determinant, of the form 
\be \sum_{i,j,k=1}^3 \varepsilon_{ijk}  \, \mu_{1 i; r} \, \mu_{2 j; s} \, \mu_{3 k; t} \ee
 satisfy $SU(3)_L \times SU(3)_R$ invariance, for any triple of integers $r,s,t=0,\ldots,N_c-1$. But we also need to impose invariance under the $U(1)_{axial}$ factor in the global symmetry $SU(3)^2 \times U(1)_{axial}$. Denoting by $r_q$ ($r_\phi$) the trial R-charge of the fundamental (adjoint) l.h.s. fields, the monopoles have R-charge
 \be R[\M^\pm]= 3(1-r_q) + (N_c-1)(1-r_\phi)- (N_c-1) \ee
 So the superpotential $\CW=\M^++\M^-$ imposes the relation
 \be 6r_q=2-(2N_c-2)\rf \ee
Using the map \ref{U3flavmap}, we deduce the trial R-charges of the surviving fields $\mu_{ij; r}$ on the magnetic side:
 \be R[\mu_{ij; r}] = 2r_q + (N_c-1-r) \cdot \rf \ee
 so for the cubic terms :
 \be  R[\mu_{i \tilde{i}; r} \mu_{j \tilde{j}; s} \mu_{k \tilde{k}; t} ] = 6 r_q + (3N_c-3-r-s-t)\rf = 2 + (N_c-1-r-s-t)\rf \ee
From these it is possible to find all the interactions with total R-charge $2$ and $U(1)_{axial}$ charge $0$: only  the cubic terms $\mu_{i \tilde{i}; r} \mu_{j \tilde{j}; s} \mu_{k \tilde{k}; t}$ with $r+s+t=N_c-1$ are present in the Wess-Zumino superpotential. 
 
 \paragraph{Confining duality.} Concluding, the duality can be written as
  \be \ba{ccc}\label{U3flavdualWZ}
\ba{c}U(N_c)\, \text{w/ adjoint} \, \phi \\
 \textrm{and $3$ flavors} \,q_i,\qt_i \\
   \CW=\M^++\M^- + \sum_{j=2}^{N_c} \b_j tr(\phi^j) \ea 
    &\Longleftrightarrow& 
\ba{c}\text{Wess-Zumino model}\\
 \text{w/ $9 N_c$ fields} \,\, \mu_{ij;r}\!\sim\!tr(\qt_i \phi^{N_c-1-r} q_j) \\
   \CW\!=\!\sum_{r,s,t}\!\sum_{i,j,k}\!\varepsilon_{ijk} 
     \mu_{1 i; r} \mu_{2 j; s} \mu_{3 k; t} \delta_{r+s+t, N_c-1} \ea 
   \ea \ee
   where $i,j,k=1,2,3$ and $r,s,t=0,\ldots,N_c-1$.
   
The  $\CZ_{S^3}$ integral identity associated to \ref{U3flavdualWZ} is the first formula of Theorem 5.6.7 of \cite{FOKKO}.

\subsection{$U(N_c)$ with $2$ flavors, $\CW=\M^+$}\label{sec:Nf=2WZ}
Now we start from the duality for $U(N_c)$ with $3$ flavors and $\CW=\M^+$, eq. \ref{U3flavdual}, and turn on a complex mass 
\be \delta\CW= m \, tr(q_3\qt_3) \qquad \Longleftrightarrow \qquad \delta\CW= m\,  \mu_{33;N_c-1}\,,\ee 
breaking the $SU(3)^2 \times U(1)^2$ global symmetry to $SU(2)^2 \times U(1)^2$.\footnote{Alternatively, we could turn on real masses in the duality \ref{U3flavdualWZ} and reach the same result.} On the r.h.s. the $\CF$-terms are satisfied by a vacuum proportional to
\be \Qt_3=(1, 0,\ldots,0) \qquad 
\Phi=\textrm{Jordan}_{N_c \times N_c }
\qquad Q_3=(0,\ldots,0,1)\ee
So the magnetic gauge group is broken completely, the low energy theory is a Wess-Zumino with model with massless fields $\mu_{ij;r}, \mu_{-;r}$ ($i,j=1,2$, $r=0,\ldots,N_c-1$). As before, the $N_c-1$ fields $\mu_{33;r}$ map to $tr(\phi^j)$, do not enter the superpotential and are free fields, so we move them on the l.h.s. adding the $\b_j$ fields.

The case $N_c=1$ duality was found in \cite{Benini:2017dud, Collinucci:2017bwv} and states the IR equivalence between $U(1)$ with $2$ flavors, $\CW=\M^+$ and a cubic Wess-Zumino with
\be  \label{WwzU12F} N_c=1\,: \qquad \CW_{WZ}= \mu_{-;0} \,det_{2\times 2}\left(\mu_{ij;0}\right) =  \mu_{-;0}\, \sum_{i,j} \varepsilon_{ij}  \, \mu_{1 i; 0} \, \mu_{2 j; 0} \ee
The analog duality for $U(N_c)$ with no adjoint and $N_f=N_c+1$ flavors, found in \cite{Benini:2017dud}, plays a crucial role in the "sequential confinement" of $3d$ quiver tails \cite{Benvenuti:2017kud}, see also \cite{Benvenuti:2017bpg,Giacomelli:2017vgk}.

We can find the quantum generated superpotential preserving the global symmetry $SU(2)^2 \times U(1)^2$ following very similar steps to the ones in section \ref{sec:Nf=3WZ}. The result is that the duality can be written as
  \be \ba{ccc}\label{U2flavdualWZ}
\ba{c}U(N_c)\, \text{w/ adjoint} \, \phi \\
 \textrm{and $2$ flavors} \,q_i,\qt_i \\
   \CW=\M^+ + \sum_{j=2}^{N_c} \b_j tr(\phi^j) \ea 
    &\Longleftrightarrow& 
\ba{c}\text{Wess-Zumino model}\\
 \text{w/ $4 N_c$ fields} \,\, \mu_{ij;r}\!\sim\!tr(\qt_i \phi^{N_c-1-r} q_j) \\
 \text{and $N_c$ fields} \,\, \mu_{-;r}\!\sim\!\M^-_{\phi^r} \\
   \CW\!=\!\sum_{r,s,t}\mu_{-;r} \sum_{i,j}\!\varepsilon_{ij}  \, \mu_{1 i; s} \, \mu_{2 j; t} \delta_{r+s+t, N_c-1} \ea 
   \ea \ee
   where $i,j=1,2$ and $r,s,t=0,\ldots,N_c-1$.
   
The  $\CZ_{S^3}$ integral identity associated to \ref{U2flavdualWZ} is the second formula of Theorem 5.6.7 of \cite{FOKKO}.

\subsection{$U(N_c)$ with $1$ flavor, $\CW=0$}\label{sec:Nf=1}
Turning on a complex mass in the duality \ref{U2flavdual}
\be \delta\CW= tr(q_2\qt_2) \qquad \Longleftrightarrow \qquad \delta\CW=  \mu_{22; N_c-1}\,,\ee 
breaks the $SU(2)^2 \times U(1)^3$ global symmetry to $U(1)^3$. On the r.h.s. the gauge symmetry $U(N_c)$ breaks completely, so the low energy theory is a Wess-Zumino model with massless fields $\mu_{11;r}, \mu_{+;r}, \mu_{-;r}$ ($r=0,\ldots,N_c-1$) and $\mu_{22;r}$ ($r=0,\ldots,N_c-2$). The $N_c-1$ fields $\mu_{22;r}$ do not enter the superpotential and are free fields, so we move them on the l.h.s. adding the $\b_j$ fields. We rename the fields $q_1, \qt_1, \mu_{11;r}\rightarrow q, \qt, \mu_{r}$.

%We now want to determine the precise superpotential of the Wess-Zumino model.
%Using the following (trial) R-charges on the electric side
%\be R[q_2,\qt_2]=1\,,\quad R[q_1,\qt_1]=r_q \quad \Rightarrow \quad R[\M^{\pm}_{\phi^j}]=1-r_q+(N_c-1)(1-2+j(1-r_\phi)) \,,\ee
%and the map \ref{U2flavmap}, we deduce the trial R-charges of the surviving fields on the magnetic side 
%\be R[\mu_{11;r}]= 2r_q +(N_c-1-r)\rf\,,\quad R[\mu_{22;r}]= 2 +(N_c-1-r)\rf \,,\quad  R[\mu_{+;r},\mu_{-;r}]= 1-r_q +(N_c-1-r)\rf\,.\ee
%From these we can make a list of the interactions with total R-charge $2$ and topological charge $0$:
%\begin{itemize}
%\item $\mu_{11; N_c-1}\mu_{+;0}\mu_{-;0}$ satisfies all the symmetries, so we expect it to be generated (in the duality  \ref{U2flavdual} such an interaction would break the $SU(2) \times SU(2)$ symmetry)
%\item Similarly for the interaction $\mu_{11; N_c-2}(\mu_{+;1}\mu_{-;0}+\mu_{+;0}\mu_{-;1})$
%\item $\ldots$
%\end{itemize}
%Putting all these term together, we conclude that a Wess-Zumino superpotential
%\be \CW_{magnetic, IR} = \sum_{r=0}^{N_c-1} \mu_{11;N_c-1-r} \sum_{s=0}^r \mu_{12;s} \mu_{21;r-s} \ee
%is generated. 

The low energy duality can be phrased as
\be \ba{ccc}\label{Unf=1duality}
\ba{c}U(N_c)\, \text{w/ adjoint} \, \phi \\
 \textrm{and $1$ flavor} \, q,\qt \\
   \CW= \sum_{j=2}^{N_c} \b_j tr(\phi^j)  \ea 
    &\Longleftrightarrow& 
\ba{c}\text{Wess-Zumino model}\\
 \text{w/ $N_c$ fields} \,\, \mu_{ r}\!\sim\!tr(\qt \phi^{N_c-1-r} q) \\
 \text{and $2 N_c$ fields} \,\, \mu_{\pm;r}\!\sim\!\M^{\pm}_{\phi^r} \\
   \CW=  \sum_{r,s,t=0}^{N_c-1} \, \mu_{r} \, \mu_{+; s} \, \mu_{-; t} \delta_{r+s+t,N_c-1}  \ea 
   \ea \ee
The case $N_c=1$ is the well known duality between $U(1)$ with $1$ flavors and the $XYZ$ model \cite{Aharony:1997bx}.  Notice that for any $N_c$ there is a $\mathbb{Z}_3$ symmetry on the r.h.s., which emerges in the IR on the gauge theory side.
 
The $\CZ_{S^3}$ identity associated to \ref{Unf=1duality} is Theorem $5.6.8$ of \cite{FOKKO}, as was pointed out in \cite{Aghaei:2017xqe}.  One new thing is that here we spell out the explicit form of the superpotential, which contains $\frac{N_c(N_c+1)}{2}$ cubic terms. 

An equivalent version of \ref{Unf=1duality} (with all singlets $\mu_r$ except one on the l.h.s., and with the topological symmetry gauged) was derived in \cite{Benvenuti:2017lle,Benvenuti:2017kud} starting from the $\CN=4$ mirror symmetry for $SU(N_c)$ $2N_c$ flavors hyper (for which the equality of the round $S^3$ partition function can be proven as a function of the $\CN=4$ mass and Fayet-Iliopoulos parameters \cite{Benvenuti:2011ga}), and using the \emph{sequential confinement} technique.
 
 Interestingly, \cite{Nieri:2018pev} recently derived the $\CZ_{S^3}$ identity starting from $5d$ $\CN=1$ gauge theories, through S-duality for pq-webs and topological string / instantons partition functions. Similar ideas have been used in \cite{Benvenuti:2016wet,Zenkevich:2017ylb}, and might be useful to derive new $3d$ $\CN=2$ dualities from $5d$.
   
Similarly to  the duality \ref{U2flavdual2}, also \ref{Unf=1duality} admits a  brane interpretation in Type IIB superstring \`a la Hanany-Witten analogous to \ref{PIC:Nf=2duality}:
%\begin{figure}
\be\label{PIC:Nf=1duality}
\begin{tikzpicture}
\draw [ultra thick, blue] (2,2.5) -- (2,-2.5);\node at (2.35,2.5) {NS};
\draw [dashed, ultra thick, gray] (0,2.5) -- (0,-2.5);\node at (0.35,2.5) {D5'};
\draw [ultra thick, blue] (-2.5,2.5) -- (-2.5,-2.5);\node at (-2.15,2.5) {NS};
\node at (-1.3, 1.3){$N_c$};
\draw [thick, red] (-2.5,1) -- (0,1);
\draw [thick, red] (-2.5,0.7) -- (0,0.7);
\draw [thick, red] (-2.5,0.4) -- (0,0.4);
\draw [thick, dashed, red] (-2.5,-0.3) -- (0,-0.3);
\draw [thick, red] (-2.5,-1) -- (0,-1);
\node at (1, 1.3){$N_c$};
\draw [thick, red] (2,0.8) -- (0,0.8);
\draw [thick, red] (2,0.5) -- (0,0.5);
\draw [thick, red] (2,0.3) -- (0,0.3);
\draw [thick, dashed, red] (2,-0.3) -- (0,-0.3);
\draw [thick, red] (2,-0.8) -- (0,-0.8);
\node at (4,0) {$\qquad\qquad\Longleftrightarrow\qquad\qquad$};
\node at (4,0.5) {S-duality};
\end{tikzpicture} 
%\begin{tikzpicture}\node at (0,2.5) {$\qquad\Longleftrightarrow\qquad$};\end{tikzpicture}
\begin{tikzpicture}
\draw [dashed, ultra thick, gray] (2,2.5) -- (2,-2.5);\node at (2.35,2.5) {D5'};
\draw [ultra thick, blue] (0,2.5) -- (0,-2.5);\node at (0.35,2.5) {NS};
\draw [dashed, ultra thick, gray] (-2.5,2.5) -- (-2.5,-2.5);\node at (-2.15,2.5) {D5'};
\node at (-1.3, 1.3){$N_c$};
\draw [thick, red] (2,1) -- (0,1);
\draw [thick, red] (2,0.7) -- (0,0.7);
\draw [thick, red] (2,0.4) -- (0,0.4);
\draw [thick, dashed, red] (2,-0.3) -- (0,-0.3);
\draw [thick, red] (2,-1) -- (0,-1);
\node at (1, 1.3){$N_c$};
\draw [thick, red] (-2.5,0.8) -- (0,0.8);
\draw [thick, red] (-2.5,0.5) -- (0,0.5);
\draw [thick, red] (-2.5,0.3) -- (0,0.3);
\draw [thick, dashed, red] (-2.5,-0.3) -- (0,-0.3);
\draw [thick, red] (-2.5,-0.8) -- (0,-0.8);
\end{tikzpicture}  \ee%\end{figure}   

\subsection{$Usp(2N_c)$ with $6$ flavors, $\CW=\M$}
We can deform the duality \ref{BASEDUAL} adding on the electric side a mass term for two of the $8$ flavors: 
\be \delta\CW_{electric}=tr(q_7q_8) \quad \Longleftrightarrow \quad \delta\CW_{magnetic}= \mu_{78; N_c-1} \ee
The global symmetry is broken $SU(8)\times U(1) \rightarrow SU(6) \times U(1)$. On the magnetic side we are turning on a linear superpotential and, as usual in Seiberg dualities, this induces a Higgsing of the gauge group. The $\CF$-terms of the fields $\mu_{ij; l}$ imply that $tr(Q_7 A^{N_c-1} Q_8)$ must take a vev, while all the other
 $tr(Q_iA^{l}Q_j)$'s stay at zero vev. This is accomplished taking the vev for the antisymmetric field $A$ to be $\left(\ba{cc} 0& 1\\1&0\ea\right) \otimes \textrm{Jordan}_{N_c \times N_c}$, where $\textrm{Jordan}_{N_c \times N_c}$ is a maximal Jordan block. The vev breaks completely the magnetic gauge group, the $12 N_c$ fields $\mu_{7j; r}$ and $\mu_{8 j; r}$ ($j=1,\ldots,6$, $r=0,\dots,N_c-1$) get a mass and disappear from the spectrum. The fields $\mu_{78; r}$ decouple, so we add the $\b_j$ fields and $\CW=\sum \b_j tr(a^j)$.\footnote{It is interesting to perform A-maximization \cite{Intriligator:2003jj, Anselmi:1997am} for the electric gauge theory in $4d$, $Usp(2N_c)$ with antisymmetric $a$ and $6$ fundamentals $q_i$, with $ \CW_{4d}=\sum_{j=2}^{N_c}\b_j tr(a^j)$. The result for the superconformal R-charges, for all $N_c$, is very simple
   \be R[a]=0 \qquad R[q_i]=\frac{1}{3} \ee
So all the electric mesons $tr( q_i a^r q_j)$ have $R=\frac{2}{3}$ and are free fields in the IR. Since the $tr( q_i a^r q_j)$ are precisely at the unitarity bound we can leave the $\mu_{ij; r}$ on the r.h.s. of the duality. Accordingly, in $4d$ the cubic Wess-Zumino interactions are marginally irrelevant. We see another example of how unitarity issues are mapped under duality to asymptotic freedom issues.

Notice that since $R[a]=0$, the $\CZ_{S^3 \times S^1}$ of the electric theory "factorizes" into a product of the $\CZ_{S^3 \times S^1}$ of $N_c$ $SU(2)$ gauge theories with $6$ doublets. This is analogous to the $SU(N_c) \leftrightarrow U(1)^{N_c-1}$ Abelianization in $3d$ \cite{Benvenuti:2017kud}.}

In the case $N_c=1$, found by Intriligator and Pouliot \cite{Intriligator:1995ne}, the Wess-Zumino superpotential is the $SU(6)$-invariant Pfaffian of the $6 \times 6$ mesonic matrix:
  \be \ba{ccc}
\ba{c}Usp(2)=SU(2)\, \text{w/ $6$ flavors} \, q_i \\
   \CW=\M \ea 
    &\Longleftrightarrow& 
\ba{c} \text{Wess-Zumino model} \\
 \textrm{w/ $15$ chiral fields} \, \mu_{ij}\sim tr(q_i q_j)  \\
   \CW=  \textrm{Pfaff}_{6 \times 6}(\mu_{ij}) = \frac{1}{2^3 3!}\sum_{ijklmn} \varepsilon_{ijklmn}\mu_{ij}\mu_{kl}\mu_{mn} \ea 
   \ea \ee
The generalization to $N_c > 1$ follows the same steps of the previous cases. The IR duality reads
  \be \label{usp6flavM}\ba{ccc}
\ba{c}Usp(2N_c)\, \text{w/ antisymmetric} \, a \\
 \textrm{and $6$ flavors} \, q_i \\
   \CW=\M+\sum_{j=2}^{N_c}\b_j tr(a^j) \ea 
    &\Longleftrightarrow& 
\ba{c} \text{Wess-Zumino model} \\
 \textrm{w/ $15 N_c$ chiral fields} \, \mu_{ij; l}\sim tr( q_i a^l q_j)  \\
   \CW\!=\!\frac{1}{2^3 3!} \sum_{rst} \! \sum_{ijklmn} \varepsilon_{ijklmn}\mu_{ij; r}\mu_{kl; s}\mu_{mn; t} \delta_{r+s+t,N_c-1} \ea 
   \ea \ee

The associated $\CZ_{S^3}$ integral identity is formula 5.3.7 of \cite{FOKKO}.

\subsection{$Usp(2N_c)$ with $4$ flavors, $\CW=0$}
Finally, we start from \ref{usp60Wflavdual}  and turn on a complex mass $tr(q_5 q_6)$, (or from \ref{usp6flavM} and turn on real masses $0,0,0,0,+1-1$) and get a Wess-Zumino dual of $Usp(2N_c)$ with $4$ flavors, $\CW=0$. The case $N_c=1$ was found by Aharony \cite{Aharony:1997gp}. With the same steps as before, we arrive at the duality
  \be \label{usp4flav0W} \ba{ccc}
\ba{c}Usp(2N_c)\, \text{w/ antisymmetric} \, a \\
 \textrm{and $4$ flavors} \, q_i. \\
   \CW= 0 \ea 
    &\Longleftrightarrow& 
\ba{c} \text{Wess-Zumino model} \\
 \text{w/ $6 N_c$ singlets} \,\, \mu_{ij;r}\sim tr( q_i a^l q_j). \\
 \text{and $N_c$ singlets} \,\, \mu_{56; r}\sim \M_{a^r}. \\
 \CW\!=\!\frac{1}{2^2 2!} \sum_{rst} \! \sum_{ijkl} \varepsilon_{ijkl} \mu_{ij; r}\mu_{kl; s}\mu_{56; t} \delta_{r+s+t,N_c-1} \ea 
   \ea \ee
   
The associated $\CZ_{S^3}$ integral identity is the first formula in Theorem 5.6.6 of \cite{FOKKO}.

%%%%%%%%%%%%%%%%%%%%%%%%%%%%%%%%%%%%%%%%%%%%%%%%%%%%%%%%%%%%%%%%%%%%%%%%%%%%%%%%%%%%%%%%%%%%%%%%%%%%%%%%%%%%%%%%%%%%%%%%%%%%%%%%%%%%%%%%%%%%%%%%%%%%%%%%%%%%%%%%%%%%%%%%%%%%%%%%%%%%%%%%%%%%%%%%
\section{Axial + complex masses: "Duality appetizers"}\label{sec:appetizers}
In this section we start from the confining dualities of section \ref{sec:confining} and turn on real masses for the flavors such that the Wess-Zumino models become free.\footnote{It would be easy to flip enough fields in the Wess-Zumino models and obtain dualities with a set of free fields (for instance in the $U(1)$ with $1$ flavor $q,\qt$ $\leftrightarrow XYZ$ duality, we can flip the meson $q\qt \leftrightarrow X$ and obtain the duality between $\CN\!=\!4$ $U(1)$ with $1$ flavors and two free chirals $Y,Z$), but this not what we are interested in here. Here we want a duality between a gauge theory with no flipping fields on the l.h.s., and a set of free fields on the r.h.s.} In this section we allow for non-vector-like (axial) real masses, so in particular Chern-Simons terms are generated.

If zero flavors are left, we recover two classes of dualities that were discovered in \cite{Kapustin:2011vz} generalizing the findings of \cite{Jafferis:2011ns} for adjoint-$SU(2)$:\footnote{To be precise, at low energies the gauge theory is not completely free, there is also a topological quantum field theory in the infrared. Describing the topological field theory in each case goes beyond the scope of this paper.}
\be\label{appetu} \quad U(N_c)_1 \,\, \text{w/ adjoint $\phi$,}\,\,\, \CW=0 \,\,\,\,\,\qquad  \Longleftrightarrow \quad N_c\!-\!1 \,\, \text{free fields} \,\, \sim tr(\phi^j)\ee
\be\label{appetusp} Usp(2N_c)_2 \,\,\text{w/ antisymm $a$,}\,\,\, \CW=0 \quad  \Longleftrightarrow \quad N_c\!-\!1 \,\, \text{free fields} \,\, \sim tr(a^j) \ee
The free fields on the r.h.s. are simply the Casimir invariant combinations $tr(a^j)$ and $tr(\phi^j)$ (recall that both $a$ and $\phi$ are traceless, $tr(a)\!=\!tr(\phi)\!=\!0$, so $j=2,\ldots,N_c$). There are no monopoles in the chiral ring because of the Chern-Simons interaction.

Let us mention that, in section \ref{sec:AD}, inspired by a $4d$ susy enhancement duality, we discover new dualities for $Usp(2N_c)_{\frac12}$ with an adjoint field (instead of an antisymmetric) and $2$ fundamentals $q,p$, from which, turning on a real mass for $q$, we get:
 \be \label{usp2flavadj} \ba{ccc}
\ba{c}Usp(2N_c)_{\frac12}\, \text{w/ adjoint} \, \phi \\
 \textrm{and $1$ flavor} \, p ,\,   \CW= tr(p \phi p) \ea 
    &\Longleftrightarrow& 
\ba{c} \text{$N_c$  free fields} \,\, \sim tr(\phi^{2j}).  \ea 
   \ea \ee

It is easy to see that the dualities \ref{appetusp} and \ref{appetu} can be obtained from the dualities \ref{usp4flav0W} and \ref{Unf=1duality} (respectively), turning on same sign real masses for all the flavors.

It is however interesting to spell out all such dualities obtained from the confining dualities of section \ref{sec:confining}. We need to turn on axial-like real masses which do not involve the $2$-index matter, so we need at least two abelian factors in the global symmetry: we cannot turn on an axial real mass for in $Usp(2N_c)$ with $6$ flavors and $\CW=\M$, or in $U(N_c)$ with $(3,3)$ flavors and $\CW=\M^++\M^-$ (the global symmetry is only $SU(6) \times U(1)$, or $SU(3)^2 \times U(1)$). So in the rest of this section we turn on axial real masses for the three dualities  \ref{U2flavdualWZ},  \ref{Unf=1duality} and \ref{usp4flav0W}, producing more examples of infrared free gauge theories.

In all the dualities of this section, the gauge theory is not time-reversal invariant in the UV (because of the Chern-Simons interactions), but the dual side is free so it is obviously time-reversal invariant. The duality then implies that time-reversal invariance in these gauge theories is emergent. In other words the statements hold also reversing the sign of the Chern-Simons term.

In this section we present the dualities leaving all the free fields on the r.h.s., as in \cite{Jafferis:2011ns} and \cite{Kapustin:2011vz}.

\subsection{$U(N_c)$ with $0$ and $(1,0)$ flavors, $\CW=0$}
We start from the duality \ref{Unf=1duality} with global symmetry $U(1)^3$
\be \ba{ccc}\label{appetU11}
\ba{c}U(N_c)\, \text{w/ adjoint} \, \phi \\
 \textrm{and $(1,1)$ flavors} \, q, \qt \\
   \CW=0  \ea 
    &\Longleftrightarrow& 
\ba{c} \text{Wess-Zumino w/ $3N_c$ fields}\, \mu_{r} \,,\mu_{\pm;r} \\
 \text{plus $N_c\!-\!1$ free fields}\, \sim tr(\phi^j) \\
   \CW=  \sum_{r=0}^{N_c-1} \mu_{N_c-1-r} \sum_{s=0}^r \mu_{+;s} \mu_{-;r-s}  \ea 
   \ea \ee
Turning on a positive real mass for $\qt$, on the l.h.s. a Chern-Simons term $\frac{1}{2}$ is generated, so the dressed monopoles $\M^-_{\phi^r}$ are not gauge invariant anymore. On the r.h.s. $\mu_r$ and $\mu_{+;r}$ become massive. \ref{appetU11} flows to 
\be \ba{ccc}\label{appetU10}
\ba{c}U(N_c)_{\frac{1}{2}}\, \text{w/ adjoint} \, \phi \\
 \textrm{and $(1,0)$ flavor} \, q \\
   \CW=0  \ea 
    &\Longleftrightarrow& 
\ba{c} \text{$N_c$ free monopoles}\,\, \sim \M^+_{\phi^r} \\
 \text{plus $N_c\!-\!1$ free fields}\, \sim tr(\phi^j)
  \ea 
   \ea \ee
The case $N_c=1$ gives rise to the most basic $\CN\!=\!2$ duality:  $U(1)_{\frac12}$ with $1$ charged chiral dual to a free chiral field, found in \cite{Dorey:1999rb}.

Turning on a positive real mass for $q$ in \ref{appetU10}, on the l.h.s. no monopole is left in the chiral ring, and on the r.h.s. the free monopoles get a mass. \ref{appetU10} flows to \ref{appetu}:
\be \ba{ccc}\label{appetU00}
\ba{c}U(N_c)_1\, \text{w/ adjoint} \, \phi \\
 \textrm{and $0$ flavors},\,   \CW=0  \ea 
    &\Longleftrightarrow& 
\ba{c}  \text{$N_c\!-\!1$ free fields}\,\sim tr(\phi^j) \ea 
   \ea \ee
The associated $\CZ_{S^3}$ integral identities are in Theorem 5.6.8 of \cite{FOKKO}.

The chain of dualities in this subsection was already discussed in \cite{Amariti:2014lla}.

Since we have a brane setup for \ref{appetU11}, we can easily produce brane setups for the \ref{appetU10} and \ref{appetU00}. First we move the $D5'$ on top of a $NS5$ so that they form a pq-web with $4$ semi-infinite $5$-branes. Adding the first real mass corresponds to deform the pq-web to the minimal pq-web with $3$ semi-infinite $5$ branes of type $(1,0), (0,1), (1,1)$. Adding the second real mass corresponds to deforming the minimal pq-web with $3$ semi-infinite $5$ branes into a single $5$-brane of type $(1,1)$. It is indeed known that $N_c$ $D3$'s streching between a $(1,0)$ and $(1,k)$ 5-brane give rise to a  $U(N_c)_k$ gauge theory.

\subsection{$U(N_c)$ with $(2,1)$ and $(2,0)$ flavors, $\CW=\M^+$}
If we turn on real masses for the antifundamentals in the confining duality for $U(N_c)$ with $(2,2)$ flavors, $\CW=\M^+$, \ref{U2flavdualWZ}, which has global symmetry $SU(2)^2 \times U(1)^2$, we get
\be \ba{ccc}\label{appetU21}
\ba{c}U(N_c)_{\frac{1}{2}}\, \text{w/ adjoint} \, \phi \\
 \textrm{and $(2,1)$ flavor} \, q_i, \qt \\
   \CW=\M^+  \ea 
    &\Longleftrightarrow& 
\ba{c} \text{$2N_c$ free mesons}\,\, \sim tr(\qt \phi^r q_i) , \\
 \text{and $N_c\!-\!1$ free fields}\, \sim tr(\phi^j)
  \ea 
   \ea \ee
   and
\be \ba{ccc}\label{appetU20}
\ba{c}U(N_c)_{1}\, \text{w/ adjoint} \, \phi \\
 \textrm{and $(2,0)$ flavor} \, q_i \\
   \CW=\M^+  \ea 
    &\Longleftrightarrow& 
\ba{c}\text{ $N_c\!-\!1$ free fields}\, \sim tr(\phi^j)  \ea 
   \ea \ee
At this point the global symmetry is only $SU(2) \times U(1)$ and we cannot turn on axial real masses for the flavors anymore.
   
The associated $\CZ_{S^3}$ integral identities are in Theorem 5.6.7 of \cite{FOKKO}.

\subsection{$Usp(2N_c)$ with $0,1,2,3$ real flavors $\CW=0$}
\cite{Kapustin:2011vz} also found the duality $Usp(2N_c)_2$ with antisymmetric $\leftrightarrow$ $N_c\!-\!1$ free chirals. 

Starting from the duality for $Usp(2N_c)$ with $4$ flavors and $\CW=0$, \ref{usp4flav0W}, with global symmetry $SU(4) \times U(1)^2$, and turning on a single real mass for $q_4$, on the l.h.s. the tower of monopoles and $3$ of the $6$ towers of mesons disappear from the chiral ring. On the r.h.s. it is impossible to write a non trivial interaction for the $\mu_{ij; r}$ (basically there is no $SU(3)_L \times SU(3)_R$ invariant made with a $3 \times 3$ antisymmetric matrix).
 \be \label{usp3flav0W} \ba{ccc}
\ba{c}Usp(2N_c)_{\frac12}\, \text{w/ antisymmetric} \, a \\
 \textrm{and $3$ flavors} \, q_i ,\,   \CW= 0 \ea 
    &\Longleftrightarrow& 
\ba{c}  \text{3$N_c$ free mesons} \,\, \sim tr( q_i a^l q_j) \\
 \text{and $N_c\!-\!1$  free fields} \,\, \sim tr(a^r).  \ea 
   \ea \ee
We then turn on a real mass for $q_3$, $q_2$ and $q_1$, flowing to 
 \be \label{usp2flav0W} \ba{ccc}
\ba{c}Usp(2N_c)_{1}\, \text{w/ antisymmetric} \, a \\
 \textrm{and $2$ flavors} \, q_i ,\,   \CW= 0 \ea 
    &\Longleftrightarrow& 
\ba{c}  \text{$N_c$ free mesons} \,\, \sim tr( q_1 a^l q_2) \\
 \text{and $N_c\!-\!1$ free fields} \,\, \sim tr(a^r) , \ea 
   \ea \ee
 \be \label{usp1flav0W} \ba{ccc}
\ba{c}Usp(2N_c)_{\frac32}\, \text{w/ antisymmetric} \, a \\
 \textrm{and $1$ flavor} \, q ,\, \CW= 0 \ea 
    &\Longleftrightarrow& 
\ba{c}   \text{$N_c\!-\!1$ free fields} \,\, \sim tr(a^r)   \ea 
   \ea \ee
and
 \be \label{usp0flav0W} \ba{ccc}
\ba{c}Usp(2N_c)_{2}\, \text{w/ antisymmetric} \, a \\
   \CW= 0 \ea 
    &\Longleftrightarrow& 
\ba{c}   \text{$N_c\!-\!1$ free fields} \,\, \sim tr(a^r) .  \ea 
   \ea \ee
The associated $\CZ_{S^3}$ integral identities are in Theorem 5.6.6 of \cite{FOKKO}.

%%%%%%%%%%%%%%%%%%%%%%%%%%%%%%%%%%%%%%%%%%%%%%%%%%%%%%%%%%%%%%%%%%%%%%%%%%%%%%%%%%%%%%%%%%%%%%%%%%%%%%%%%%%%%%%%%%%%%%%%%%%%%%%%%%%%%%%%%%%%%%%%%%%%%%%%%%%%%%%%%%%%%%%%%%%%%%%%%%%%%%%%%%%%%%%%
\section{Axial masses $1$: $U(N_c) \leftrightarrow Usp(2N_c)$}\label{sec:axial1}   
In this section we describe a descending chain of dualities relating $U(N_c)$ and $Usp(2N_c)$ gauge theories. 
%The $Usp$ theories have $f$ fundamentals and Chern-Simons level $3-\frac{f}{2}$, with $f=6,5,4,3,2,1,0$. 
Each line in the following scheme represents a duality
 \be\label{UUspscheme}
\begin{tikzpicture}
\node (a) at (0,4.5) {$U(N_c)_{\CW=\M^++\M^-}$ w/$(4,4)$ flav};
\node (b) at (0,3) {$U(N_c)_{\CW=\M^+}$ w/$(3,3)$ flav $$};
\node (c) at (0,1.5) {$U(N_c)$ w/$(2,2)$ flav};
\node (d) at (0,0) {$U(N_c)_{-\frac12}$ w/$(2,1)$ flav};
\node (e) at (-2.5,-1.5) {$U(N_c)_{-1}$ w/$(2,0)$ flav};
\node (f) at (2.5,-1.5) {$U(N_c)_{-1}$ w/$(1,1)$ flav};
\node (g) at (0,-3) {$U(N_c)_{-\frac32}$ w/$(1,0)$ flav};
\node (h) at (0,-4.5) {$U(N_c)_{-2}$ w/$0$ flav};
\draw [->,ultra thick, red] (a) -- (b);%\node at (3.3,1.5) {\tiny{$+,+,+,-,-,-$}};
\draw [->, ultra thick, red] (b) -- (c);%\node at (-3,1.5) {\tiny{$0,0,0,0,+,-$}};
\draw [->,ultra thick, green] (c) -- (d);%\node at (3.3,-1.5) {\tiny{$0,0,+;0,0,-$}};
\draw [->, ultra thick, green] (d) -- (e);%\node at (-2.8,-1.5) {\tiny{$+,+,-,-$}};
\draw [->,ultra thick, green] (d) -- (f);%\node at (1,-4.5) {\tiny{$0,+;0,-$}};
\draw [->, ultra thick, green] (e) -- (g);%\node at (-2.8,-1.5) {\tiny{$+,+,-,-$}};
\draw [->,ultra thick, green] (f) -- (g);%\node at (1,-4.5) {\tiny{$0,+;0,-$}};
\draw [->, ultra thick, green] (g) -- (h);%\node at (-2.8,-1.5) {\tiny{$+,+,-,-$}};
\node (a1) at (9,4.5) {$Usp(2N_c)$ w/$3\!+\!\bar{3}$ flav};
\node (b1) at (9,3) {$Usp(2N_c)_{\frac12}$ w/$3\!+\!\bar{2}$ flav};
\node (c1) at (9,1.5) {$Usp(2N_c)_{1}$ w/$2\!+\!\bar{2}$ flav};
\node (d1) at (9,0) {$Usp(2N_c)_{\frac32}$ w/$2\!+\!\bar{1}$ flav};
\node (e1) at (9,-1.5) {$Usp(2N_c)_{2}$ w/$2$ flav};
\node (f1) at (9,-3) {$Usp(2N_c)_{\frac52}$ w/$1$ flav};
\node (g1) at (9,-4.5) {$Usp(2N_c)_{3}$ w/$0$ flav};
\draw [->,ultra thick] (a1) -- (b1);%\node at (3.3,1.5) {\tiny{$+,+,+,-,-,-$}};
\draw [->, ultra thick] (b1) -- (c1);%\node at (-3,1.5) {\tiny{$0,0,0,0,+,-$}};
\draw [->,ultra thick] (c1) -- (d1);%\node at (3.3,-1.5) {\tiny{$0,0,+;0,0,-$}};
\draw [->, ultra thick] (d1) -- (e1);%\node at (-2.8,-1.5) {\tiny{$+,+,-,-$}};
\draw [->,ultra thick] (e1) -- (f1);%\node at (1,-4.5) {\tiny{$0,+;0,-$}};
\draw [->, ultra thick] (f1) -- (g1);%\node at (-2.8,-1.5) {\tiny{$+,+,-,-$}};
\end{tikzpicture}\ee
Except for the first two flows, from the $U(N_c)$ perspective, a single axial mass is turned on at each step (it is easy to read from the flavors which disappear). We use red and blue for RG flows triggered by vector-like real masses, and green color for RG flows triggered by axial real masses.

We choose to put all the gauge singlets on the $Usp$ side, but of course they can be moved on the $U(N_c)$ side, providing instances of $U(N_c)$ theories with enhanced global symmetries.

Let us mention that if we turn on a complex mass in the dualities of the scheme \ref{UUspscheme}, we flow to a confining duality only for the cases with no Chern-Simons, otherwise we flow to a "duality appetizer" of section \ref{sec:appetizers}.

The set of  $Usp$ theories on the right of \ref{UUspscheme} satisfies self-dualities of form
\be \label{SDusp}Usp(2N_c)_k \,\, \text{with $6\!-\!2k$ flavors} \Longleftrightarrow Usp(2N_c)_{-k} \,\, \text{with $6\!-\!2k$ flavors}\ee
coming from either the \emph{Csaki et al. like} (only the "baryons" are flipped) or the \emph{Intriligator-Pouliot like} duality (both the "baryons" and the "mesons" are flipped) for $Usp(2N_c)$ with $6$ flavors. For $N_c=1$ they reduce to the dualities $SU(2)_k \leftrightarrow SU(2)_{-k}$ with $6\!-\!2k$ doublets \cite{Willett:2011gp}.

Similarly, one can reduce the \emph{Aharony like} and \emph{mirror like} self-dualities studied in section \ref{sec:U2flavoW} of $U(N_c)$ with $2$ flavors and obtain self-dualities for the Chern-Simons $U(N_c)$ models on the left of \ref{UUspscheme}.

In order to get $U(N_c) \leftrightarrow Usp(2N_c)$ dualities, we start from the \emph{Csaki et al. like} duality for $Usp(2N_c)$ with $8$ flavors \ref{BASEDUALcsaki} and give masses $(+, +, +, -; -, -, -, +)$ to the electric $q_i$ flavors.

On the electric side there are $4$ positive and $4$ negative masses, so the same arguments leading to \ref{U4flavdual} apply, and we are left with $U(N_c)$ with an adjoint $\phi$ and $(4,4)$ fundamentals $q_i, \qt_i$. However, on the magnetic side the dual vacuum is the origin of the moduli space, that does not break the $Usp(2N_c)$ gauge group. The flavors $Q_{1,2,3}, \Qt_{1, 2, 3}$ and the $6$ towers of gauge singlets $b_{i4; r}$ and $\bt_{i4; r}$ ($i=1,2,3$) are massless, while the flavors $Q_{4}, \Qt_{4}$  and the singlets $b_{ij; r}$ and $\bt_{ij; r}$ ($i<j=1,2,3$) become massive.

In the IR we get the following $U(N_c) \leftrightarrow Usp(2N_c)$ duality with $3+3$ towers of gauge singlets $b_{ij;r}$ and $\bt_{ij;r}$, flipping the "baryons" of the $6=3\!+\!\bar{3}$ split:
  \be \ba{ccc}\label{U44Usp33f}
\ba{c}U(N_c)\, \text{w/ adjoint $\phi$}\\ \text{and $(4,4)$ flavors}  \, q_i,\qt_j \\
   \CW=\M^+ + \M^- \ea 
    &\Longleftrightarrow& 
\ba{c}Usp(2N_c)\, \text{w/ antisymm $A$}\\ \text{ and $3\!+\!\bar{3}$ flavors} \, Q_i, \Qt_j \\
   \CW= \sum_{r=0}^{N_c-1}\sum_{i,j,k=1}^3 \varepsilon_{ijk} b_{k 4; r}tr(Q_i \Phi^{r} Q_j)+ \\
   \quad +\sum_{r=0}^{N_c-1}\sum_{i,j,k=1}^3 \varepsilon_{ijk} \bt_{k 4; r}tr(\Qt_i \Phi^{r} \Qt_j)   \ea 
   \ea \ee
The global symmetry is $SU(4)^2 \times U(1)$. On the r.h.s. the UV global symmetry is $SU(3)^2 \times U(1)^3$. 
The $\CZ_{S^3}$ integral identity associated with this duality is Theorem 5.6.15 of \cite{FOKKO}. The chiral ring generators map as
\be \ba{cccc}\label{mapU44Usp33}
 \left\{\ba{c}
 tr( \qt_i \phi^r q_j) \,,\,\, tr( \qt_{i} \phi^r q_{4}) \\  
 tr( \qt_{4} \phi^r q_j) \,,\,\, tr( \qt_{4} \phi^r q_{4})  
  \\ tr( \phi^l ) \ea  \right\}
 &\Longleftrightarrow&
   \left\{ \ba{c} tr( \Qt_i A^r Q_j) \,,\quad   b_{i4; N_c-r-1}\\ 
                    \bt_{4j; N_c-r-1} \,,\,\,  \M_{A^l}  \\
                      tr( A^l )  \ea  \right\} &
   \ba{c} i,j=1,2,3 \\ j=1,2,3 \\ l=2,\ldots,N_c\ea
   \ea \ee
Notice the mapping meson $\rightarrow$ monopole. Throughout this section, on the $Usp$ side, the singlets flip all the "baryons" of the split.

In \ref{U44Usp33f} we can only turn on vector-like real masses. Turning on $+m$ for $q_1$ and $-m$ for $\qt_4$ on the l.h.s, one of the monopole superpotentials gets lifted. On the r.h.s. only $Q_1$ gets a mass, the monopole disappears from the chiral ring, and the $2$ towers of singlets $\bt_{24; r}$, $\bt_{34; r}$ become massive. At the end we flow to
  \be \ba{ccc}\label{U3fUsp5f}
\ba{c}U(N_c)\, \text{w/ adjoint $\phi$}\\ \text{ and $(3,3)$ flavors}  \, q_i,\qt_j \\
   \CW=\M^+  \ea 
    &\Longleftrightarrow& 
\ba{c}Usp(2N_c)_{\frac12}\, \text{w/ antisymm $A$,}\\ \text{ $3\!+\!\bar{2}$ flavors} \, Q_i, \Qt_j \, \text{and $4N_c$ singlets}\\
   \CW= \sum_{r=0}^{N_c-1}\sum_{i,j,k=1}^3 \varepsilon_{ijk} b_{k 4; r}tr(Q_i \Phi^{r} Q_j)+ \\
   \quad +\sum_{r=0}^{N_c-1}  \bt_{12;r} tr(\Qt_1 A^r \Qt_2) \ea 
   \ea \ee
The $\CZ_{S^3}$ integral identity associated with this duality is Theorem 5.6.16 of \cite{FOKKO}. The chiral ring generators map as 
\be \ba{cccc}\label{mapU33Usp32}
 \left\{\ba{c}
 tr( \qt_i \phi^r q_j) \,,\,\, tr( \qt_{i} \phi^r q_{3}) \\  
% tr( \qt_{3} \phi^r q_j) \,,\,\, tr( \qt_{3} \phi^r q_{3})  \\
  \M^-_{\phi^r}  \\ tr( \phi^l ) \ea  \right\}
 &\Longleftrightarrow&
   \left\{ \ba{c} tr( Q_i A^r \Qt_j) \,,\quad   b_{i3; N_c-r-1}\\ 
    %               tr( \Qt_3 A^r Q_j)  \,,\,\,  b_{23; N_c-r-1} \\
        \bt_{12; N_c-r-1}       \\      tr( A^l )  \ea  \right\} &
   \ba{c} i=1,2,3 , \, j=1,2  \\  \\ l=2,\ldots,N_c\ea
   \ea \ee

At this point the global symmetry is $SU(3)^2 \times U(1)^2$, so we have a choice: we can turn on another vector-like real mass, or an axial mass. We explore the second option in section \ref{sec:axial2}. Turning on a vector-like real mass $(0,0,+1;-1,0,0)$ in \ref{U3fUsp5f} we flow to a duality studied (for generic $N_c$) in \cite{Amariti:2014lla}:
  \be \ba{ccc}\label{U2fUsp4f}
\ba{c}U(N_c)\, \text{w/ adjoint $\phi$}\\ \text{ and $(2,2)$ flavors}  \, q_i,\qt_j \\
   \CW=0  \ea 
    &\Longleftrightarrow& 
\ba{c}Usp(2N_c)_{1}\, \text{w/ antisymm $A$,}\\ \text{$2\!+\!\bar{2}$ flavors} \, Q_i, \Qt_j\\
\text{and $2N_c$ singlets}  \, b_{12;r},\bt_{12;r} \,\sim \M^\pm_{\phi^{N_c-1-r}}\\
   \CW= \sum_{r=0}^{N_c-1}  b_{12;r} tr(Q_1 A^r Q_2)+\\
   +\sum_{r=0}^{N_c-1}  \bt_{12;r} tr(\Qt_1 A^r \Qt_2) \ea 
   \ea \ee
The $\CZ_{S^3}$ integral identity associated with this duality is 2nd formula in Theorem 5.6.17 of \cite{FOKKO}. In \ref{U2fUsp4f} (and in all the dualities that follow in this section), $U(N_c)$ mesons map to $Usp(2N_c)$ "mesons", $U(N_c)$ monopoles map to singlets, that flip the $Usp(2N_c)$ "baryons".

The case $N_c=1$ of \ref{U2fUsp4f} (composed with Aharony duality of $U(1)$ with $2$ flavors or with the \emph{Intriligator-Pouliot like} \ref{SDusp}), is equivalent to a duality appeared in \cite{Aharony:2014uya}.\footnote{We need to set $F=\tilde{F}=N=2, k=1$ in the duality of \cite{Aharony:2014uya}, that is
  \be \ba{ccc}\label{AF}
\ba{c} SU(N)_{k}\, \text{w/ $(F,\tilde F)$ flavors} \\
   \CW= 0 \ea 
    &\Longleftrightarrow& 
\ba{c}U\left(\frac{F+\tilde{F}}{2} - N + k\right)_{-k, \frac{F+\tilde{F}}{2} - N }\\ \text{w/ $(F,\tilde F)$ flavors}  \, q_i,\qt_j \\
   \CW= \sum_{ij} \mu_{ij} tr(\qt_i q_j)  \ea 
   \ea \ee
Using the \emph{Intriligator-Pouliot like} \ref{SDusp} dualities for the $SU(2)$ side, which also reverses the Chern-Simons level, \ref{AF} for $N=2$ becomes:
   \be \ba{ccc}\label{AF2}
\ba{c} SU(2)_{-k}\, \text{w/ $(F,\tilde F)$ flavors} \\
   \CW= \text{"baryons flipped"} \ea 
    &\Longleftrightarrow& 
\ba{c}U\left(\frac{F+\tilde{F}}{2} - 2 + k\right)_{-k, \frac{F+\tilde{F}}{2} - 2 }\\ \text{w/ $(F,\tilde F)$ flavors} , \, 
   \CW= 0  \ea 
   \ea \ee
Also all the next dualities in this section, for $N_c=1$, are special cases of \ref{AF2}.} Moving the singlets on the l.h.s. we get a $U(N_c)$ gauge theory with IR enhanced global symmetry $SU(2)^2 \times U(1)^3\rightarrow SU(4) \times U(1)^2$. Such symmetry enhancement can also be proven composing the  \emph{Aharony like} duality \ref{U2flavdual} with the \emph{mirror like} duality \ref{U2flavdual2}: the arguments are the same as in \cite{Benini:2018bhk}, simply replacing $U(1) \rightarrow U(N_c)$ and the gauge singlets with a tower of $N_c$ gauge singlets.

 Turning on masses $(0,0;0,1)$ in \ref{U2fUsp4f} we flow to
  \be \ba{ccc}\label{U21fUsp3f}
\ba{c}U(N_c)_{\frac12}\, \text{w/ adjoint $\phi$}\\ \text{ and $(2,1)$ flavors}  \, q_i,\qt \\
   \CW=0  \ea 
    &\Longleftrightarrow& 
\ba{c}Usp(2N_c)_{\frac32}\, \text{w/ antisymm $A$,}\\ \text{$2\!+\!\bar{1}$ flavors} \, Q_i, \Qt \\
\text{and $N_c$ singlets} \, b_{12; r} \,\sim \M^+_{\phi^{N_c-1-r}} \\
   \CW = \sum_{r=0}^{N_c-1} b_{12; r} tr(Q_1 A^r Q_2) \ea 
   \ea \ee
 (2nd formula of Theorem 5.6.18 of \cite{FOKKO}.)
 
From \ref{U21fUsp3f} we have two options: we can flow to a duality with no singlets (already discussed in \cite{Amariti:2014lla})
  \be \ba{ccc}\label{U11fUsp2f}
\ba{c}U(N_c)_{1}\, \text{w/ adjoint $\phi$}\\ \text{ and $(1,1)$ flavors}  \, q,\qt \\
   \CW=0  \ea 
    &\Longleftrightarrow& 
\ba{c}Usp(2N_c)_{2}\, \text{w/ antisymm $A$}\\ \text{ and $1\!+\!\bar{1}$ flavors} \, Q, \Qt \\
   \CW= 0 \ea 
   \ea \ee
   (2nd formula of Theorem 5.6.19 of \cite{FOKKO}, notice the IR symmetry enhancement in the $U(N_c)$ gauge theory) or to a duality with $1$ tower of singlets:
  \be \ba{ccc}\label{U20fUsp2f}
\ba{c}U(N_c)_{1}\, \text{w/ adjoint $\phi$}\\ \text{ and $(2,0)$ flavors}  \, q_i \\
   \CW=0  \ea 
    &\Longleftrightarrow& 
\ba{c}Usp(2N_c)_{2}\, \text{w/ antisymm $A$}\\ \text{$2$ flavors} \, Q_1, Q_2 \\
\text{and $N_c$ singlets} \, \mu_r \sim \M^+_{\phi^r} \\
   \CW= \sum_{r=0}^{N_c-1} b_{12; r} tr(Q_1 A^r Q_2)\ea 
   \ea \ee
(1st formula of Theorem 5.6.19 of \cite{FOKKO}.)

 From both \ref{U11fUsp2f} and \ref{U20fUsp2f}, we flow to 
  \be \ba{ccc}\label{U1fUsp1f}
\ba{c}U(N_c)_{\frac32}\, \text{w/ adjoint $\phi$}\\ \text{ and $(1,0)$ flavor} ,\,
   \CW=0  \ea 
    &\Longleftrightarrow& 
\ba{c}Usp(2N_c)_{\frac52}\, \text{w/ antisymm $A$}\\ \text{ and $1$ flavor}  ,\,
   \CW= 0 \ea 
   \ea \ee
   (Theorem 5.6.21 of \cite{FOKKO}), and finally to
   \be \ba{ccc}\label{U0fUsp0f}
\ba{c}U(N_c)_{2}\, \text{w/ adjoint $\phi$} ,\,   \CW=0  \ea 
    &\Longleftrightarrow& 
\ba{c}Usp(2N_c)_{3}\, \text{w/ antisymm $A$} ,\,   \CW= 0 \ea 
   \ea \ee
  The $\CZ_{S^3}$ integral identity associated with this duality is Theorem 5.6.22 of \cite{FOKKO}, which is the \underline{last} theorem of the Thesis.

\subsection{One chiral flavor and supersymmetry enhancement}\label{susyenh}
\cite{Gang:2018huc} recently proposed that the $\CN=2$ $U(1)_{\frac32}$ theory with  $1$ chiral flavor (with full UV global symmetry $U(1)_{top} \times U(1)_R$) displays supersymmetry enhancement to $\CN=4$ supersymmetry. Highly non trivial evidence was given in support of the statement \cite{Gang:2018huc}. 

The $U(1)_{\frac32}$ theory is the case $N_c=1$ of the l.h.s. of the duality \ref{U1fUsp1f}.\footnote{We are grateful to Francesco Benini for suggestions and discussions about this subsection.} 

Notice that $\CN=2$ $SU(2)_k$ theories with precisely one complex doublet $q^\a$ have UV global symmetry $SU(2)_F$: in $\CN=1$ language, the superpotential (including the real adjoint $\CN=1$ superfield) is
\be \CW_{\CN=1} = \sum_{\a,\b=1,2} (S^{\a\b} q_a q^\dagger_\b - \frac{k}{2} S^{\a\b}S_{\a\b} )\ee
Integrating out the massive $SU(2)$-adjoint $S^{\a\b}$, one gets a quartic superpotential, and only one structure can appear
\be \CW_{\CN=1} \sim \left(\sum_{\a} q^\a q^\dagger_\a \right)^2 \sim \left(\sum_{A,B,I,J} q_{A, I} q_{B, J} \varepsilon^{AB} \varepsilon^{I J}\right)^2 \ee
The last expression is the only quartic $SU(2)_F$-invariant written doubling the components of $q^{\a}$ to $q_{A, I}$ and imposing a reality condition $q_{A, I} = \varepsilon^{AB} \varepsilon^{IJ} q^*_{B, J}$, where $A,B$ are gauge indices, and $I,J$ are $SU(2)_F$ indices. So \ref{U1fUsp1f} "explains" part of the IR symmetry enhancement to $SO(4)_R$. What should happen is that only if $k=\pm \frac52$ also the $U(1)$ R-symmetry of $SU(2)_k$ with one doublet enhances the IR $U(1)_R \rightarrow SU(2)$, and the full IR global symmetry is $SO(4)_R$, which is the R-symmetry of $\CN=4$ supersymmetry.

Another curious facts about the theories $U(N_c)_{\frac32} \leftrightarrow Usp(2N_c)_\frac52$ appearing in  the duality \ref{U1fUsp1f} is that, since the $Usp \leftrightarrow Usp$ dualities \ref{SDusp} relate $Usp(2N_c)_\frac52$ with $Usp(2N_c)_{-\frac52}$ (with just one flavor there are no gauge invariants operators to flip), the theories $U(N_c)_{\frac32} \leftrightarrow Usp(2N_c)_\frac52$ with one flavor are time-reversal invariant in the IR.

These facts are consistent with the natural conjecture that \emph{for any $N_c$ the following theories have enhanced $\CN=4$ supersymmetry:}
\be U(N_c)_{\pm \frac32} \text{w/ adjoint and $1$ chiral flavor} \Longleftrightarrow Usp(2N_c)_{\pm \frac52}\, \text{w/ antisymm and $1$ flavor} \ee
We leave a more detailed investigation of this proposal to future work.

%%%%%%%%%%%%%%%%%%%%%%%%%%%%%%%%%%%%%%%%%%%%%%%%%%%%%%%%%%%%%%%%%%%%%%%%%%%%%%%%%%%%%%%%%%%%%%%%%%%%%%%%%%%%%%%%%%%%%%%%%%%%%%%%%%%%%%%%%%%%%%%%%%%%%%%%%%%%%%%%%%%%%%%%%%%%%%%%%%%%%%%%%%%%%%%%
\section{Axial masses $2$:  $U(N_c)_{\CW=\M^+}$ $\leftrightarrow$ $U(N_c)_{\CW=0}$}\label{sec:axial2}   
In this section we start from duality \ref{U3fUsp5f}, and turn on an axial mass for $q_1$. On l.h.s. a Chern-Simons term is generated, and the monopoles $\M^-$ exit the chiral ring. On the r.h.s. only one fundamental and the singlets $\bt_{12; r}$ become massive, so we flow to:
  \be \ba{ccc}\label{U32fUsp4f}
\ba{c}U(N_c)_\frac12\, \text{w/ adjoint $\phi$}\\ \text{ and $(3,2)$ flavors}  \, q_i,\qt_j \\
   \CW=\M^+  \ea 
    &\Longleftrightarrow& 
\ba{c}Usp(2N_c)_{1}\, \text{w/ antisymm $A$}\\ \text{and $3\!+\!\bar{1}$ flavors} \, Q_i, \Qt \,,\\
%\text{and $3N_c$ singlets}  \\
   \CW= \sum_{r=0}^{N_c-1}\sum_{i,j,k=1}^3  \varepsilon_{ijk} b_{k;r} tr(Q_i A^r Q_j) \ea 
   \ea \ee
   The chiral ring generators map as 
\be \ba{cccc}\label{mapU32Usp31}
 \left\{\ba{c}
 tr( \qt_1 \phi^r q_i) \,,\,\, tr( \qt_{2} \phi^r q_{i}) \\   tr( \phi^l ) \ea  \right\}
 &\Longleftrightarrow&
   \left\{ \ba{c} tr( Q_i A^r \Qt) \,,\quad   b_{i; N_c-r-1}\\             tr( A^l )  \ea  \right\} &
   \ba{c} i=1,2,3  \\  l=2,\ldots,N_c\ea
   \ea \ee
Combining \ref{U32fUsp4f} and \ref{U2fUsp4f} (after moving the singlets on the $U(N_c)$ side using the chiral ring maps), we get a $U\leftrightarrow U$ duality:
   \be \ba{ccc}\label{U32flavdual22}
\ba{c}U(N_c)_{\frac12}\, \text{w/ adjoint} \, \phi \\
 \textrm{and $(3,1\!+\!\bar{1})$ flavors} \,q_i,\qt_i \\
   \CW=\M^+ + \sum_r \sum_{i=1}^3 \mu_{i;r} tr(q_i \phi^r \qt_2) \ea 
    &\Longleftrightarrow& 
\ba{c}U(N_c)\, \text{w/ adjoint} \, \Phi \\
 \textrm{and $(2,1\!+\!\bar{1})$ flavors} \, Q_i, \Qt_i \\
   \CW= \sum_{r=0}^{N_c-1}\sum_{\pm} \mu_{\pm;r} \M^\pm_{\Phi^r} \ea 
   \ea \ee
The chiral ring generators map as
\be \ba{cccc}\label{mapU32Usp31}
 \left\{\ba{c}
 tr( \qt_1 \phi^r q_i) \,,\,\, \mu_{i; r} \\ 
  tr( \qt_1 \phi^r q_3) \,,\,\, \mu_{3; r} \\   tr( \phi^l ) \ea  \right\}
 &\Longleftrightarrow&
   \left\{ \ba{c} tr( Q_i A^r \Qt_1) \,,\quad tr( Q_i A^r \Qt_2) \\      
    \mu^+_{N_c-1-r} \,,\quad   \mu^-_{N_c-1-r} \\      tr( A^l )  \ea  \right\} &
   \ba{c} i=1,2 \\ \\  l=2,\ldots,N_c\ea
   \ea \ee
As can be seen from the above mapping, a curious phenomenon is at work here: the l.h.s. UV global symmetry includes an $SU(3) \times U(1)$,  the r.h.s. UV global symmetry includes an $SU(2)_L \times SU(2)_R \times U(1)_{top}$. While $SU(2)_L$ is a subgroup of $SU(3)$, $SU(2)_R$ does not commute with $SU(3)$. In the IR SCFT the $SU(3) \times U(1)$ and the $SU(2)^2 \times U(1)_{top}$ combine into an $SU(4)$, which is explicit in the $Usp$ duality frame. Both sides have symmetry enhancement to an IR global symmetry is $SU(4) \times U(1)^2$. 

Now we turn on an axial real mass for $\qt_2$, flowing to 
   \be \ba{ccc}\label{U32flavdual22}
\ba{c}U(N_c)_{1}\, \text{w/ adjoint} \, \phi \\
 \textrm{and $(3,1)$ flavors} \,q_i,\qt \\
   \CW=\M^+ + \sum_r \sum_{i=1}^3 \mu_{i;r} tr(q_i \phi^r \qt) \ea 
    &\Longleftrightarrow& 
\ba{c}U(N_c)_{-\frac12}\, \text{w/ adjoint} \, \Phi \\
 \textrm{and $(2,1)$ flavors} \, Q_i, \Qt \\
   \CW= \sum_{r=0}^{N_c-1} \mu_{+;r} \M^+_{\Phi^r} \ea 
   \ea \ee
The $3$ singlets $\mu_{i;r}$ map to $2$ mesons and $\mu_{+;r}$.

Finally we turn on an axial real mass for $\qt$, flowing to 
   \be \ba{ccc}\label{U32flavdual22}
\ba{c}U(N_c)_{\frac32}\, \text{w/ adjoint} \, \phi \\
 \textrm{and $(3,0)$ flavors} \,q_i \\
   \CW=\M^+  \ea 
    &\Longleftrightarrow& 
\ba{c}U(N_c)\, \text{w/ adjoint} \, \Phi \\
 \textrm{and $(2,0)$ flavors} \, Q_i, \Qt \\
   \CW=0 \ea 
   \ea \ee
%This chiral ring is trivial in this case. 
Notice the emerging time reversal symmetry on the l.h.s. and the emerging $SU(3)$ on the r.h.s. 

The case $N_c=1$ appeared recently in \cite{Fazzi:2018rkr}, which pointed out that it explains the symmetry enhancement $SU(2) \times U(1)_{top} \rightarrow SU(3)$ discovered in \cite{Gang:2017lsr, Gang:2018wek} and studied \emph{via} an $\CN=1$ Wess-Zumino model in \cite{Gaiotto:2018yjh,Benini:2018bhk}.
%%%%%%%%%%%%%%%%%%%%%%%%%%%%%%%%%%%%%%%%%%%%%%%%%%%%%%%%%%%%%%%%%%%%%%%%%%%%%%%%%%%%%%%%%%%%%%%%%%%%%%%%%%%%%%%%%%%%%%%%%%%%%%%%%%%%%%%%%%%%%%%%%%%%%%%%%%%%%%%%%%%%%%%%%%%%%%%%%%%%%%%%%%%%%%%%
\section{The mirror of $A_{2N}$ Argyres-Douglas and adjoint-$Usp$ dualities}\label{sec:AD}   

As pointed out in \cite{Aghaei:2017xqe}, starting from the $4d$ $Usp \leftrightarrow Usp$ dualities, one can flow to the duality \ref{Unf=1duality}. Upon gauging the $U(1)$ symmetry, one gets the equivalent duality for the $SU(N_c)$ theory:  
\be \label{SUUduality} \ba{ccc}
\ba{c}SU(N_c)\, \text{w/ adjoint} \, \phi \\
 \textrm{and $1$ flavor} \,  q,\qt \,.\\
   \CW=  \sum_{j=2}^{N_c} \b_j tr(\phi^j) \ea 
    &\Longleftrightarrow& 
\ba{c} U(1)\, \textrm{w/} \, N_c \, \textrm{flavors} \, \mu_{\pm; r} \,.\\
   \CW=  \sum_{r=0}^{N_c-1} \mu_{N_c-1-r} \sum_{s=0}^r \mu_{+;s} \mu_{-;r-s}  \ea 
   \ea \ee
where the $N_c$ dressed mesons $tr(\qt \phi^r q)$ map to singlets $\mu_r$, the $2$ dressed baryons map to monopoles and the $N_c(N_c-1)$ dressed monopoles of $SU(N_c)$ map to mesons $ \mu_{+;s} \mu_{-;t}$. See \cite{Benvenuti:2017kud} for details about the map.
 It is important that for the dualities \ref{SUUduality} and \ref{Unf=1duality} the UV global symmetry is $U(1)^3$ on both sides. Accordingly, the $\CZ_{S^3}$'s match as a function of $3$ fugacities.
 
We can "move" the singlets $\mu_{r}$ on the l.h.s. (i.e. flip them with fields $\a_{r}$, recalling that $tr(\qt \phi^r q) \leftrightarrow \mu_{r}$), then the global symmetry enhances. Starting from \ref{SUUduality} and moving $N_c-1$ singlets, we get the duality
  \be \label{SUUdualN=4} \ba{ccc}
\ba{c}SU(N_c)\, \text{w/ adjoint} \, \phi \\
 \textrm{and $1$ flavor} \,  q,\qt \,.\\
   \CW=  \sum_{r=0}^{N_c-2} \a_r tr(\qt \phi^r q) + \\+ \sum_{j=2}^{N_c} \b_j tr(\phi^j) \ea 
    &\Longleftrightarrow& 
\ba{c} U(1)\, \textrm{w/} \, N_c \, \textrm{flavors} \, \mu_{\pm; r} \,.\\
   \CW= \mu_{N_c-1} \sum_{s=0}^{N_c} \mu_{+; s} \mu_{-; N_c-s}  \\
   \CN=4 \, \text{susy} \ea 
   \ea \ee
This is precisely the duality studied in \cite{Benvenuti:2017lle, Benvenuti:2017kud}, and obtained using "sequential confinement" from a $\CN\!=\!4$ mirror symmetry for $U(N_c)$ with $2N_c$ flavors. The r.h.s. of \ref{SUUduality}, however, displays $\CN\!=\!4$ supersymmetry and non-R global symmetry $U(1)_{top} \times SU(N_c)$. This symmetry enhancement is obvious on the r.h.s., but the Cartan generators of the new symmetries only emerge in the IR on the l.h.s., making it impossible to match the $S^3$ partition functions as functions of all the UV fugacities visible on the r.h.s. So in a sense the presentation of the duality \ref{SUUduality} is more natural than the presentation \ref{SUUdualN=4}.

The l.h.s. of \ref{SUUduality} is the $3d$ reduction of the theory proposed in \cite{Maruyoshi:2016tqk,Maruyoshi:2016aim}, modified as in  \cite{Benvenuti:2017lle, Benvenuti:2017kud} in order to have a consistent superpotential that does not violate the \emph{chiral ring stability} criterion of  \cite{Benvenuti:2017lle}. The r.h.s. theory, $U(1)$ with $N_c$ flavors and $\CN=4$ supersymmetry is expected to be the mirror dual of $A_{2N_c-1}$ Argyres-Douglas models compactified to $3d$ \cite{Nanopoulos:2010bv,Xie:2012hs}. See \cite{Buican:2015ina, Buican:2015hsa, Agarwal:2017roi, Benvenuti:2017bpg, Evtikhiev:2017heo, Giacomelli:2017ckh, Giacomelli:2018ziv} for additional results about $\CN=1$ lagrangians for $4d$ $\CN=2$ Argyres-Douglas theories, and \cite{Benini:2018bhk,Gaiotto:2018yjh,Gang:2018huc, Buican:2018ddk} for other recent examples of supersymmetry enhancement.

\paragraph{$Usp(2N_c)$ with adjoint and $2$ flavors, $\CW=tr(p \phi p)$: $A_{2N_c}$  AD models.}
We now study the analogous case of $A_{2N_c}$ Argyres-Douglas.

The $4d$ $A_{2N_c}$ Argyres-Douglas theory, as far as we know, does not have a proposed $3d$ mirror.\footnote{We are grateful to Mathew Buican for discussions about this.} It is however known that $4d$ Higgs Branch is just a point. Since Higgs Branches are not expected to change compactifying theories with $8$ supercharges, we assume that the Higgs Branch is trivial also in $3d$. This means that the Coulomb Branch of the mirror must be trivial. Assuming that the $3d$ mirror is an $\CN=4$ theory with an $\CN=4$ Lagrangian, the only possibility is that the $3d$ mirror has no gauge interactions, so it must be free. Since the $4d$ Coulomb Branch has rank-$N_c$, we reach the conclusion

\hspace{1cm}\emph{the $3d$ mirror of $A_{2N_c}$ AD is $N_c$ free hypermultiplets.}

Maruyoshi and Song \cite{Maruyoshi:2016tqk,Maruyoshi:2016aim} proposed that a certain $4d$ $Usp(2N_c)$ with adjoint and $\phi$ and $2$ flavors $q,p$ flows in the IR to $A_{2N_c}$ AD:
\be\label{A2ndual4d}\ba{c}Usp(2N_c)\,  \text{with adjoint} \, \phi\\  \text{and $2$ flavors} \, p , q ,\\
 \CW_{4d}= tr(p\phi p) + \sum_{r=1}^{N_c} \a_r tr(q \phi^{2r-1} q) +\\+\sum_{j=0}^{N_c-1}  \b_j tr(\phi^{2j+2})  \ea
  \Longrightarrow \ba{c} (A_1,A_{2N_c}) \, \text{Argyres-Douglas theory}\ea \ee
Using the prescription of \cite{Benvenuti:2017lle, Benvenuti:2017kud}, we added the $N_c$ singlets $\b_j$ to deal with the (apparent) unitarity violations for the operators $tr(\phi^{2j})$.
The UV global symmetry in $4d$ is just $U(1)$, so one performs A-maximization in one variable, the result is:
\be\label{rel1} R[q]=1-\frac{4 N+3}{3 (2 N+3)} \qquad \qquad R[p]=1 - \frac{1}{3 (3 + 2 N)}\qquad\qquad R[\phi]= \frac{2}{3 (2 N+3)} \ee
The operators $\a_r$ and $\b_r$ have R-charges
\be\label{rel2} R[\a_r] =\frac{2}{3} \frac{4 N+4-2 r}{2 N+ 3} \qquad\qquad R[\b_{r}] = \frac{2}{3}\frac{6N + 7 - 2 r}{2 N+3} \ee
The operators $\a_r$ map to the Coulomb Branch generators of $A_{2N_c}$ Argyres-Douglas \cite{Maruyoshi:2016tqk,Maruyoshi:2016aim}.

From \ref{rel1} and \ref{rel2}, we notice that the following $N_c$ relations hold:
\be R[\b_r]=R[\a_r]+\frac23 \qquad\qquad R[tr(pq)]=R[\a_{N_c}]+\frac23\ee
The interpretation of these relations is that the $N_c$ holomorphic operators $tr(pq)$ and $\b_r$ ($r=2,\ldots,N_c$) are not in the chiral ring, but sit inside the $\CN=2$ supersymmetric Coulomb Branch multiplet: they are descendant (under the emerging $\CN=2$ supersymmetry) of the $\a_r$'s ($r=1,\ldots, N_c$). See \cite{Benvenuti:2017lle, Benvenuti:2017kud} for the analog discussion in the $A_{2N-1}$ case. It is not clear to us what happens to $\b_0$ in this case (it is easy to see that $\b_0$ cannot take a vev, so it is likely not in the chiral ring).

Now we flow to $3d$. No monopole superpotential is generated, as for the $SU(N_c)$ case in \cite{Benvenuti:2017lle, Benvenuti:2017kud}. Combining with the $3d$ mirror proposed above, we are then led to conjecture the following duality
\be\label{A2ndual}\ba{c}Usp(2N_c)\,  \text{with adjoint} \, \phi\\  \text{and $2$ flavors} \, p , q ,\\
 \CW= tr(p\phi p) + \sum_{r=1}^{N_c} \a_r tr(q \phi^{2r-1} q) +\\+\sum_{j=1}^{N_c}  \b_j tr(\phi^{2j})  \ea
  \Longleftrightarrow \ba{c} 2N_c\, \text{free chiral fields} \\ \qquad \sim Usp(2N_c)\text{-monopoles} \ea \ee

The free fields on the r.h.s. are the monopoles, indeed, for a $Usp(2N_c)$ with adjoint, we can construct $3N_c$ $\CN\!=\!4$ Coulomb Branch operators \cite{Cremonesi:2013lqa}: the Casimirs of the adjoint $tr(\phi^{2j})$, $j=1,\ldots,N_c$, and $2N_c$ BPS dressed monopoles $\M_{\phi^j}$, $j=0,1,\ldots,2N_c-1$. The UV global symmetry on the l.h.s. is $U(1)^2$, both $U(1)$ factors mix with R-symmetry.

We checked numerically the duality \ref{A2ndual}. Performing $\CZ$-extremization, we find
\be r_q=\frac12 , \qquad \rf=0 , \qquad r_p=1\,. \ee

Recalling that 
\be R[\M] = 1-r_q + 1-r_p + 2N_c(1-\rf)+2N_c(1-2) = 1-r_q - \left(2N_c-\frac12\right)\rf \ee
(the last contribution comes from the gauginos), we see that the superconformal R-charge of all the dressed monopoles is $\frac12$, so they are free fields. All the $N_c$ $\a_r$ fields in $3d$ have R-charge $1$, so they must be a quadratic combination of the monopoles. These are quantum relations that are non-trivial to see in the gauge theory. It is easy to see that the global symmetries allow for the quantum relations, for instance if $N_c=1$, $R[\a_1]=2-2r_q-\rf$ and $R[\M]=1-r_q - \frac32 \rf$, so a chiral ring relation
\be \a_1 = \M_\phi^2 \ee
is consistent with the global symmetries.

%$3d$ dualities with non-trivial quantum relations have been studied in \cite{Giacomelli:2017vgk}.

It would be nice to uplift the duality above to a $3d$ confining duality and possibly to a $3d$ or $4d$ duality for $Usp$ with adjoint, flavors and cubic "dressed meson" superpotential. We leave these interesting issues to future work.

%%%%%%%%%%%%%%%%%%%%%%%%%%%%%%%%%%%%%%%%%%%%%%%%%%%%%%%%%%%%%%%%%%%%%%%%%%%%%%%%%%%%%%%%%%%%%%%%%%%%%%%%%%%%%%%%%%%%%%%%%%%%%%%%%%%%%%%%%%%%%%%%%%%%%%%%%%%%%%%%%%%%%%%%%%%%%%%%%%%%%%%%%%%%%%%%%%%%%%%%%%%%%%%%
\section{A "mirror" dual of $U(N_c)$ with adjoint and $N_f$ flavors, $\CW=0$}\label{sec:mirror}
In this section we consider a generalization of the duality discussed in section \ref{sec:Nf=1}. We look for a dual of $\CN\!=\!2$ adjoint-$U(N_c)$ with $N_f$ flavors, with arbitrary $N_f$. Let us read off the duality from the generalized brane setup. On the l.h.s. we stretch $N_c$ D3's between two NS5's, and add $N_f$ flavor D5's branes. As in figure \ref{PIC:Nf=1duality}, the $NS5$ branes stretch along $012345$, the $D3$ branes along $0126$ and the $D5'$ branes along $012347$. The central $D5'$ on the l.h.s. provides the flavor $q_1,\qt_1$ to the $U(N_c)$ gauge theory.

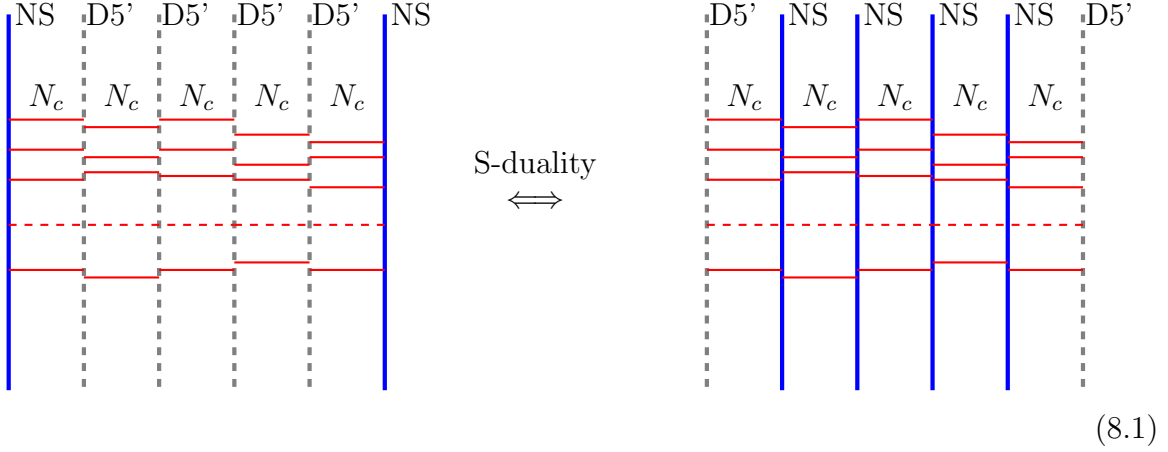
\be\label{branesNF}
\begin{tikzpicture}
\draw [ultra thick, blue] (-3,2.5) -- (-3,-2.5);\node at (-2.65,2.5) {NS};
\draw [dashed, ultra thick, gray] (-2,2.5) -- (-2,-2.5);\node at (-2+0.35,2.5) {D5'};
\draw [dashed, ultra thick, gray] (-1,2.5) -- (-1,-2.5);\node at (-1+0.35,2.5) {D5'};
\draw [dashed, ultra thick, gray] (0,2.5) -- (0,-2.5);\node at (0.35,2.5) {D5'};
\draw [dashed, ultra thick, gray] (1,2.5) -- (1,-2.5);\node at (1.35,2.5) {D5'};
\draw [ultra thick, blue] (2,2.5) -- (2,-2.5);\node at (2.35,2.5) {NS};
\node at (-2.5, 1.4){$N_c$};\node at (-1.5, 1.4){$N_c$};\node at (-0.5, 1.4){$N_c$};\node at (0.5, 1.4){$N_c$};\node at (1.5, 1.4){$N_c$};
%\node at (1, 1.3){$N_c$};
\draw [thick,red] (-3,1.1)--(-2,1.1);\draw [thick, red] (-2,1)--(-1,1);\draw[thick, red] (-1,1.1) -- (0,1.1);\draw [thick, red] (0,0.9) -- (1,0.9);\draw [thick, red] (1,0.8) -- (2,0.8);
\draw [thick, red] (-3,0.7) -- (-2,0.7);\draw [thick, red] (-2,0.6) -- (-1,0.6);\draw [thick, red] (-1,0.7) -- (0,0.7);\draw [thick, red] (0,0.5) -- (1,0.5);\draw [thick, red] (1,0.6) -- (2,0.6);
\draw [thick, red] (-3,0.3) -- (-2,0.3);\draw [thick, red] (-2,0.4) -- (-1,0.4);\draw [thick, red] (-1,0.35) -- (0,0.35);\draw [thick, red] (0,0.3) -- (1,0.3);\draw [thick, red] (1,0.2) -- (2,0.2);
\draw [thick, dashed, red] (-3,-0.3) -- (0,-0.3);\draw [thick, dashed, red] (2,-0.3) -- (0,-0.3);
\draw [thick, red] (-3,-0.9) -- (-2,-0.9);\draw [thick, red] (-2,-1) -- (-1,-1);\draw [thick, red] (-1,-0.9) -- (0,-0.9);\draw [thick, red] (0,-0.8) -- (1,-0.8);\draw [thick, red] (1,-0.9) -- (2,-0.9);
\node at (4,0) {$\qquad\qquad\Longleftrightarrow\qquad\qquad$};
\node at (4,0.5) {S-duality};
\end{tikzpicture} 
%\begin{tikzpicture}\node at (0,2.5) {$\qquad\Longleftrightarrow\qquad$};\end{tikzpicture}
\begin{tikzpicture}
\draw [dashed, ultra thick, gray] (-3,2.5) -- (-3,-2.5);\node at (-3+0.35,2.5) {D5'};
\draw [ultra thick, blue] (-2,2.5) -- (-2,-2.5);\node at (-2+0.35,2.5) {NS};
\draw [ultra thick, blue] (-1,2.5) -- (-1,-2.5);\node at (-1+0.35,2.5) {NS};
\draw [ultra thick, blue] (0,2.5) -- (0,-2.5);\node at (0.35,2.5) {NS};
\draw [ultra thick, blue] (1,2.5) -- (1,-2.5);\node at (1.35,2.5) {NS};
\draw [dashed, ultra thick, gray] (2,2.5) -- (2,-2.5);\node at (2.35,2.5) {D5'};
\node at (-2.5, 1.4){$N_c$};\node at (-1.5, 1.4){$N_c$};\node at (-0.5, 1.4){$N_c$};\node at (0.5, 1.4){$N_c$};\node at (1.5, 1.4){$N_c$};
%\node at (1, 1.3){$N_c$};
\draw [thick,red] (-3,1.1)--(-2,1.1);\draw [thick, red] (-2,1)--(-1,1);\draw[thick, red] (-1,1.1) -- (0,1.1);\draw [thick, red] (0,0.9) -- (1,0.9);\draw [thick, red] (1,0.8) -- (2,0.8);
\draw [thick, red] (-3,0.7) -- (-2,0.7);\draw [thick, red] (-2,0.6) -- (-1,0.6);\draw [thick, red] (-1,0.7) -- (0,0.7);\draw [thick, red] (0,0.5) -- (1,0.5);\draw [thick, red] (1,0.6) -- (2,0.6);
\draw [thick, red] (-3,0.3) -- (-2,0.3);\draw [thick, red] (-2,0.4) -- (-1,0.4);\draw [thick, red] (-1,0.35) -- (0,0.35);\draw [thick, red] (0,0.3) -- (1,0.3);\draw [thick, red] (1,0.2) -- (2,0.2);
\draw [thick, dashed, red] (-3,-0.3) -- (0,-0.3);\draw [thick, dashed, red] (2,-0.3) -- (0,-0.3);
\draw [thick, red] (-3,-0.9) -- (-2,-0.9);\draw [thick, red] (-2,-1) -- (-1,-1);\draw [thick, red] (-1,-0.9) -- (0,-0.9);\draw [thick, red] (0,-0.8) -- (1,-0.8);\draw [thick, red] (1,-0.9) -- (2,-0.9);
\end{tikzpicture}  \ee%\end{figure}   
The theory on the l.h.s. is adjoint $U(N_c)$ with $N_f$ flavors in the fundamental, $N_f$ flavors in the anti-fundamental (corresponding to strings stretching from the D3 to the D5' branes). The superpotential vanishes. Notice that the adjoint field is an $SU(N_c)$-adjoint plus a decoupled singlet.

The theory on r.h.s. is not so simple to read off. The theory is a quiver $U(N_c)^{N_f-1}$, with $N_f-2$ bifundamentals $b_J, \bt_J$. Each node has an adjoint field $\Phi_J$. These fields enter the superpotential in an $\CN\!=\!4$ fashion, i.e. with cubic terms of the form $tr(b_J \Phi_J \bt_J - b_J \Phi_{J-1} \bt_J )$. At this point is each adjoint field $\Phi_J$ contains also a $U(N_c)$-singlet field, which enters the superpotential.

  \be\label{QUIVER} \bpic  \path  (-4.5,0) node[rectangle,draw](x0) {$\,1\,$} -- (-2.5,0) node[circle,draw](x1) {$\,N_c\,$} -- (-0.5,0) node[circle,draw](x2) {$N_c$} -- (1.5,0) node(x3) {$\cdots$} -- (3.5,0)  node[circle,draw](x5) {$N_c$} -- (5.5,0) node[rectangle,draw](x6) {$\,1\,$} ;
\draw [->] (x0) to[bend left] (x1); \draw [<-] (x0) to[bend right] (x1);
\draw [->] (x1) to[bend left] (x2); \draw [<-] (x1) to[bend right] (x2);
\draw [->] (x2) to[bend left] (x3); \draw [<-] (x2) to[bend right] (x3);
\draw [->] (x3) to[bend left] (x5); \draw [<-] (x3) to[bend right] (x5);
\draw [->] (x5) to[bend left] (x6); \draw [<-] (x5) to[bend right] (x6);
\draw [->] (x1) to[out=-45, in=0] (-2.5,-1) to[out=180,in=225] (x1);
\draw [->] (x2) to[out=-45, in=0] (-0.5,-1) to[out=180,in=225] (x2);
\draw [->] (x5) to[out=-45, in=0] (3.5,-1) to[out=180,in=225] (x5);
\node[below right] at (-2.4,-0.6){$\Phi_1$};
\node[below right] at (-0.4,-0.6){$\Phi_2$};
\node[below right] at (3.6,-0.6){$\Phi_{N_f-1}$};
\node[above right] at (-1.8,0.3){$b_1$};
\node[above right] at (0.2,0.3){$b_2$};
\node[above right] at (2.0,0.3){$b_{N_f-2}$};
\node[below right] at (-1.7,-0.3){$\bt_1$};
\node[below right] at (0.3,-0.3){$\bt_2$};
\node[below right] at (2.0,-0.3){$\bt_{N_f-2}$};
\node[above right] at (-4,0.3){$q_L$};\node[below right] at (-3.9,-0.3){$\qt_L$};
\node[above right] at (4.2,0.3){$q_R$};\node[below right] at (4.3,-0.3){$\qt_R$};
  \epic\ee 

The external D5' branes provide a single flavor, $q_L, \qt_L$ and $q_R,\qt_R$, for each of the two external gauge nodes. This flavor enters the superpotential in a non trivial way. There are also two towers of chiral gauge-singlet fields $\a_{L,r}, \a_{R,r}$, $r=0,\ldots,N_c-1$ (these complex $\a$ modes are associated to the motion of the leftmost and rightmost D3 branes along the $34$ directions). The gauge-singlets couple to the "dressed mesons", providing an $\CN\!=\!2$ part in the superpotential:
\be \delta \CW_{\CN\!=\!2} = \sum_{r=0}^{N_c-1} \a_{L; r} tr(\qt_L \Phi_1^k q_L) +  \sum_{r=0}^{N_c-1} \a_{R; r} tr(\qt_R \Phi_{N_f-1}^k q_R) \ee

Out of the $N_f-1$ singlets in the adjoint fields $\Phi$, only $N_f-2$ linear combinations (which we call $\eta_J$) actually couple with the rest of theory. The decoupled one match with the decoupled singlet on the $U(N_c)$ side. So from now on all the adjoint fields are traceless.

The duality with the full superpotential on the quiver side is stated as:
  \be \label{ADJUNDUAL} \ba{ccc}
\ba{c}U(N_c)\, \text{w/ adjoint} \, \phi \\
 \textrm{and} \, N_f  \, \textrm{flavors}\, q_i,\qt_j \\
   \CW=0  \ea 
    &\qquad\Longleftrightarrow \qquad& 
\ba{c} U(N_c)^{N_f-1} \, \textrm{quiver \ref{QUIVER}} \\
%  \, [ 1 ]-U(N_c)-\ldots-U(N_c)-[ 1 ] \\
    \CW=  \sum_{J=1}^{N_f-2} \eta_J tr(\bt_J b_J ) +\\
    +\sum_{r=0}^{N_c-1} \left( \a_{L; r} tr(\qt_L \Phi_1^r q_L) +  \a_{R; r} tr(\qt_R \Phi_{N_f-1}^r q_R) \right) +\\
    + \sum_{J=1}^{N_f-2} \left(  tr(\bt_J \Phi_{J} b_J) + tr(\bt_{J} \Phi_{J+1} b_J) \right)   \ea 
   \ea \ee

\subsubsection*{The case $N_c=1$}
In this case the duality \ref{ADJUNDUAL} becomes the well known $\CN\!=\!2$ abelian mirror symmetry (with no adjoint fields): 
  \be \label{U1mirror} \ba{ccc}
\ba{c}U(1)\, \text{with} \, N_f  \, \textrm{flavors}\, q_i,\qt_j \\
   \CW=0  \ea 
    &\qquad\Longleftrightarrow \qquad& 
\ba{c} U(1)^{N_f-1} \, \textrm{quiver \ref{QUIVER}} \\
%  \, [ 1 ]-U(N_c)-\ldots-U(N_c)-[ 1 ] \\
    \CW=  \sum_{J=1}^{N_f-2} \eta_J tr(\bt_J b_J ) +\\
    + \a_{L} \qt_L q_L +  \a_{R} \qt_R q_R    \ea 
   \ea \ee

The global symmetry on the l.h.s. is 
\be \text{l.h.s.}\,: \qquad SU(N_f)^2 \times U(1)_q \times  U(1)_{top} \ee
where $U(1)_q$ is an axial symmetry under which $q_i$ and $\qt_i$ have charge $+1$, and $U(1)_{top}$ is the magnetic (or topological) symmetry.

As for the UV global symmetry on the r.h.s., the are $2N_f$ charged fields $p_i, \pt_i$
\be \text{r.h.s.}\,: \qquad  U(1)_{top}^{N_f-1} \times \prod_{i=1}^{N_f} U(1)_{axial;i } \times U(1)_b \ee
Where $U(1)_b$ is the non gauged combination of vector-like symmetries ($q_L,b_i,q_R$ have charge $+1$ and $\qt_L,\bt_i,\qt_R$ have charge $-1$) and $U(1)_{axial; i}$ gives charge $+1$ to $b_i$ and $\bt_i$.

The rank of the UV global symmetries is $2N_f$ on both sides. In the IR the r.h.s. global symmetry $U(1)_{top}^{N_f-1} \times \prod_{i=1}^{N_f} U(1)_{axial;i }$ enhances $SU(N_f)^2 \times U(1)$. It is well known that the $S^3$ partition functions match exactly as a function of $2N_f$ parameters.

The chiral ring generators (we only illustrate the case $N_f=3$) map as
\be \ba{ccc}\label{mirrormapNc1}
 \left\{\ba{c}  tr(\qt_1 q_1) , \, tr(\qt_1 q_2) , \, tr(\qt_1 q_3) \\  
 tr(\qt_2 q_1) , \, tr(\qt_2  q_2) , \, tr(\qt_2  q_3) \\ 
 tr(\qt_3 q_1) , \, tr(\qt_3  q_2) , \, tr(\qt_3  q_3) \\  
  \M^+ , \,\, \M^- \ea  \right\}  
        &\Longleftrightarrow&  
  \left\{\ba{c} \quad \a_{L} \,\,,\,\,\,\,\, \M^{+1,0} , \, \,\,\M^{+1,+1} \\
    \M^{-1,0} \, ,\,\,\,\, \eta_1 \,\, , \,\,\,\, \M^{0,+1} \\
   \M^{-1,-1} ,\,\, \M^{0,-1} , \, \a_R \\
    q_L b_1 q_R , \, \,\qt_L \bt_1 \qt_R  \ea  \right\}
     \ea \ee
The diagonal mesons map to gauge singlets, the off-diagonal mesons map to monopoles, and monopoles map to 'long mesons' in the quiver.

\subsubsection*{The case $N_f=2$}
In this case \ref{ADJUNDUAL} becomes the duality between two $U(N_c)$ theories \ref{U2flavdual2}. Notice that the duality is different from the duality \ref{U2flavdual}: \ref{ADJUNDUAL} is a generalization of the Abelian mirror symmetry and involve $4$ towers of gauge singlet flipping fields, while \ref{U2flavdual} is a generalization of the Aharony duality for $U(1)$ with $2$ flavors, and involves $6$ towers of gauge singlet flipping fields.

The global symmetry on the l.h.s. is 
\be \text{l.h.s.}\,: \qquad SU(2)^2 \times U(1)_q \times U(1)_\phi \times U(1)_{top} \ee
As for the UV global symmetry on the r.h.s., the are $4$ fields $p_i, \pt_i$ plus $1$ adjoint field, there are no independent relations coming from the superpotential, se we have
\be \text{r.h.s.}\,: \qquad  U(1)_{top} \times U(1)_{v} \times U(1)_{q_L} \times U(1)_{q_R} \times U(1)_b \ee
So the rank of the UV global symmetry is $5$ on both sides.

Let us recall the map of the chiral ring generators:
\be \ba{ccc}\label{mirrormapNf2}
 \left\{\ba{c}  tr(\qt_1 \phi^r q_1) ,  tr(\qt_1 \phi^r q_2)\\
 tr(\qt_1 \phi^r q_2) ,  tr(\qt_2 \phi^r q_2) \\ \M^+_{\phi^r} , \M^-_{\phi^r} \\ tr(\phi^r) \ea  \right\}
       &\Longleftrightarrow&  
  \left\{\ba{c}  \a_{L;N_c-1-r} \, , \,\, \M^{+}_{\Phi^r} \\ \M^-_{\Phi^r} \, , \,\, \a_{R;N_c-1-r} \\
     tr(\qt_L  q_R) , tr(\qt_R \qt_L)  \\ tr(\Phi^r) \ea  \right\}
     \ea \ee

\subsubsection*{The general case: $N_f>2$ and $N_c>1$}
The general case involves some subtleties. Let us start from the UV global symmetry. %Similar issues were encountered in \cite{Giacomelli:2017vgk}, studying the mirror dual of $\CN\!=\!2$ $SU(N)$ and $U(N)$ theories with $N_f$ flavors (without the adjoint field).

The global symmetry on the l.h.s. is 
\be \text{l.h.s.}\,: \qquad SU(N_f)^2 \times U(1)_q \times U(1)_\phi \times U(1)_{top} \ee
As for the r.h.s., the are $2N_f$ fields $p_i, \pt_i$ plus $N_f-1$ adjoint fields, subject to $N_f-1$ (if $N_f>2$ and $N_c>1$) independent relations coming from the superpotential terms $\sum_{J}tr(p_J \Phi_J \pt_J + p_{J+1} \Phi_J \pt_{J+1})$, so the UV global symmetry is 
\be \text{r.h.s.}\,: \qquad  U(1)_{top}^{N_f-1} \times U(1)_{v} \times U(1)_{q_L} \times U(1)_{q_R} \times U(1)_b \ee
We see that there is a mismatch between the rank of the UV global symmetry: $2N_f+1$ vs $N_f+3$, on the quiver side the rank is smaller. This is similar to what happens for the mirror of $U(N_c)$ with $N_f>2$ flavors and no adjoint, studied in  \cite{Giacomelli:2017vgk}.

We propose that the chiral ring generators map in analogy to \ref{mirrormapNc1} and \ref{mirrormapNf2}:
\begin{itemize}
\item  the first and last diagonal dressed mesons map to gauge singlets $\a_{L,R;j}$
\item  "central" diagonal mesons map to gauge singlets $\eta_j$ 
\item  "central" dressed diagonal mesons map to traceless linear combinations of $tr(\Phi_i^r)$
\item off-diagonal dressed mesons map to dressed monopoles of the form $\M^{0,..,0,1,..,1,0,..,0}$
\item dressed monopoles map to dressed 'long mesons'
\item the Casimirs of the adjoint map to a symmetric linear combination of $tr(\Phi_i^r)$ \end{itemize}
More in details, (again we only display the case $N_f=3$), the map is
\be \ba{ccc}
 \left\{\ba{c}  tr(\qt_1 \phi^r q_1) ,  tr(\qt_1 \phi^r q_2) , tr(\qt_1 \phi^r q_3)\\  
 tr(\qt_1 \phi^r q_2) ,  tr(\qt_2 \phi^r q_2) , tr(\qt_2 \phi^r q_3)\\
 tr(\qt_3 \phi^r q_2) ,  tr(\qt_3 \phi^r q_2) , tr(\qt_3 \phi^r q_3)\\ 
 \M^+_{\phi^r} , \M^-_{\phi^r} \\ 
 tr(\phi^r) \ea  \right\}
       &\Longleftrightarrow&  
  \left\{\ba{c}  \a_{L;N_c-1-r} \, , \,\, \M^{+1,0}_{\Phi_{1}^r}\,\,, \,\, \M^{+1,+1}_{\Phi_{1,2}^r} \\ 
 \M^{-1,0}_{\Phi_{1}^r} \,\,, \,\,\, \ldots \,\, , \,\,\,\, \M^{0,+1}_{\Phi_{2}^r} \\ 
   \M^{-1,-1}_{\Phi_{1,2}^r}  \, , \,\,  \M^{0,-1}_{\Phi_{2}^r}\, ,\,\, \a_{R;N_c-1-r} \\
     \sum_s tr(q_L \Phi_1^{s} b_{1} \Phi_2^{r-s} q_L) , \sum_s tr(\qt_R \Phi_{1}^{s} \bt_1 \Phi_{2}^{r-s} \qt_R)  \\
      \sum_i tr(\Phi_i^r) \ea  \right\}
     \ea \ee
Where the $\ldots$ stand for $\eta_1$ if $r=0$, for $tr(\Phi_1^r)\!-\!tr(\Phi_2^r)$ if $r>0$.

Let us emphasize one difference with respect to the case without adjoint field studied in \cite{Giacomelli:2017vgk}. Without the adjoint field, on the l.h.s. the are no dressed monopoles and no dressed mesons. But the dual is still a quiver similar to the one we are studying here, locally $\CN=4$ with adjoints. So on the dual quiver side, the chiral ring operators come in towers of dressed operators (mesons and monopoles). In the case without adjoint there must be quantum relations on the quiver side,\footnote{We are grateful to Simone Giacomelli for discussions about this.} which are highly non-trivial to see without knowing the duality. We see that adding matter, hence in a sense "uplifting" the duality can help with some of the issues.

It would be nice to uplift the duality \ref{ADJUNDUAL} to a duality where the ranks of the UV global symmetry is the same. One option might be to look for a dual of the r.h.s. quiver where the cubic "$\CN=4$" superpotentials are absent.

\subsection{Complex deformations}
To finish, we check the picture proposed above turning on superpotential terms.

\paragraph{Mass deformations.}
A complex mass for the "last" meson $tr(\qt_{N_f} q_{N_f}) \leftrightarrow \a_{R; N_c-1}$ induces on the quiver side a maximal nilpotent vev for $\Phi_{N_f-1}$. Also $q_L$ and $\qt_L$ take a vev. So the rightmost gauge group in the quiver is Higgsed $U(N_c) \rightarrow U(0)$. The quiver becomes one node shorter. This is consistent with the duality.

A complex mass for a "central" meson $tr(\qt_{i+1} q_{i+1}) \leftrightarrow \eta_i$ induces on the quiver side a vev for the operator $tr(b_i \bt_i)$. This implies that $b_i, \bt_i$ take a vev, which Higgses the two gauge groups adjacent to $b_i$ to the diagonal subgroup: $U(N_c) \times U(N_c) \rightarrow U(N_c)$. Again the quiver loses one node, consistently with the duality.

\paragraph{Cubic deformations.}
A more complicated operation is turning on cubic superpotentials, in the form of a meson dressed by one factor of $\phi$. On the l.h.s. we flow to $U(N_c)$ with adjoint and $N_f$ flavors and a cubic superpotential. No field disappears.

In the brane setup \ref{branesNF} this operation, on the l.h.s. rotates a $D5'$ into a $D5$ brane, so on the r.h.s. it rotates an $NS5$ into a $NS5'$ brane. ($NS5$/$NS5'$ branes stretch along $345$/$589$, $D5$/$D5'$ branes stretch along $789$/$347$).

Let us consider turning on the "last" once-dressed meson 
\be \label{cubicdef} tr(\qt_{N_f} \phi q_{N_f}) \Longleftrightarrow \a_{R; N_c-2} \ee
 On the quiver side of the duality, the $\CF$-terms of $\a_{R; N_c-2}$ induce the following vev:
\be \qt_{R}=(1,0,\ldots,0) \qquad 
\Phi_{N_c \times N_c}=\left(\ba{cc} \textrm{Jordan}_{N_c-1 \times N_c-1} & 0 \\ 0 & 0\ea \right)
\qquad q_{R}= - ( 0,\ldots,0, 1, 0 ) \ee
while all other fields stay at zero vev. This vev breaks the rightmost gauge group $U(N_c) \rightarrow U(1)$. This is consistent with the r.h.s. brane setup, since the rightmost $NS$ brane now is a $NS5'$, so, because of the S-rule, only one $D3$ brane can stretch between it and the right $D5'$ brane, the other $N_c-1$ $D3$'s now attach to the rightmost $NS5$, and the rightmost gauge group is only a $U(1)$.

Let us also consider turning on a "central" once-dressed meson, which, according to our proposal, map as
\be \label{cubicdef} tr(\qt_{i} \phi q_{i}) \Longleftrightarrow tr(\Phi_{i-1}^2)-tr(\Phi_{i}^2) \ee
On the quiver side, the two adjoint fields $\Phi_{i-1}$ and $\Phi_i$ become massive, integrating them out creates a quartic superpotential coupling involving the bifundamental stretching between the two gauge groups that lose the adjoints $\delta \CW= tr(b_{i-1}\bt_{i-1}b_{i-1}\bt_{i-1})$. Again, this result is confirmed on the brane picture: now on the r.h.s. we are rotating a middle $NS5$ into an $NS5'$, this makes the two adjacent adjoints massive, since the $D3$'s attached to the $NS5'$ cannot slide up and down anymore.

\acknowledgments{%We are grateful to  for useful discussions. 
We are grateful to Antonio Amariti, Francesco Benini, Marco Fazzi, Simone Giacomelli and Sara Pasquetti for useful discussion. We also thank Antonio Amariti and Luca Cassia for coordinating the submission of their paper with ours.
This work is supported in part by the MIUR-SIR grant RBSI1471GJ ``Quantum Field Theories at Strong Coupling: Exact Computations and Applications". S.B. is partly supported by the INFN Research Projects GAST and ST$\&$FI. Part of this project was completed at the workshop "Supersymmetric Quantum Field Theories in the Non-perturbative Regime" at the Galileo Galilei Institute for Theoretical Physics.}

\appendix

%%%%%%%%%%%%%%%%%%%%%%%%%%%%%%%%%%%%%%%%%%%%%%%%%%%%%%%%%%%%%%%%%%%%%%%%%%%%%%%%%%%%%%%%%%%%%%%%%%%%%%%%%%%%%%%%%%%%%%%%%%%%%%%%%%%%%%%%%%%%%%%%%%%

\bibliographystyle{ytphys}
%\small\baselineskip=.9\baselineskip

\end{document}